\documentclass{aa}

\usepackage{newtxtext,newtxmath}

\usepackage[T1]{fontenc}
\usepackage{ae,aecompl}

\usepackage{natbib}
\bibpunct{(}{)}{;}{a}{}{,} 

\usepackage{graphicx}   
\usepackage{amsmath}    
\usepackage{amssymb}    
\usepackage{xcolor}
\usepackage[normalem]{ulem}
\usepackage{subcaption}
\usepackage{booktabs}

\providecommand{\gaia}{\textsl{Gaia }}
\providecommand{\gaianospace}{\textit{Gaia}}
\providecommand{\kms}{km$~$s$^{-1}$ }
\providecommand{\kmsnospace}{km$~$s$^{-1}$}
\providecommand{\degr}{$^{\circ}~$}
\providecommand{\degrnospace}{$^{\circ}$}

\providecommand{\msun}{$M_\odot$ }
\providecommand{\vlos}{$V_{\text{los}}$ }

\providecommand{\sourceidnospace}{\texttt{source\_id}}

\begin{document}

\title{Vertical structure and kinematics of the LMC disc from \textsl{SDSS}/\textsl{Gaia}}

\author{Ó. Jiménez-Arranz\inst{1,2}
   \and D. Horta\inst{3}
   \and R.~P.~van der Marel\inst{4,5}
   \and D. Nidever\inst{6}
   \and C. F. P. Laporte\inst{7,2,8}
   \and E. Patel\inst{9}
   \and H.-W. Rix\inst{10}
}

\institute{{Lund Observatory, Division of Astrophysics, Lund University, Box 43, SE-221 00 Lund, Sweden}
\and
{Institut de Ciències del Cosmos (ICCUB), Universitat de Barcelona, Martí i Franquès 1, 08028 Barcelona, Spain}
\and
{Centre for Computational Astrophysics, Flatiron Institute, 162 5th Ave., New York, NY 10010, USA}
\and
{Space Telescope Science Institute, 3700 San Martin Drive, Baltimore, MD 21218, USA}
\and
{William H. Miller III Department of Physics and Astronomy, Johns Hopkins University, Baltimore, MD 21218, USA}
\and
{Montana State University, P.O. Box 173840, Bozeman, MT, USA}
\and
{LIRA, Observatoire de Paris, PSL Research University, CNRS, Place Jules Janssen, 92195 Meudon, France}
\and
{Kavli Institute for the Physics and Mathematics of the Universe (WPI), The University of Tokyo Institutes for Advanced Study, The University of Tokyo, Kashiwa, Chiba 277-8583, Japan}
\and
{Department of Physics and Astronomy, University of Utah, 115 South 1400 East, Salt Lake City, Utah 84112, USA}
\and
{Max-Planck-Institut für Astronomie, Königstuhl 17, D-69117 Heidelberg, Germany}
}

\date{Received <date> / Accepted <date>}

\abstract 
{Studies of the LMC's internal kinematics have provided a detailed view of its structure, largely thanks to the exquisite proper motion data supplied by the \gaia mission. However, line-of-sight (LoS) velocities, the third component of the stellar motion, are only available for a small subset of the current \gaia data, limiting studies of the kinematics perpendicular to the LMC disc plane.}
{We synergise new \textsl{SDSS}-IV/V LoS velocity measurements with existing \gaia DR3 data, increasing the 5D phase-space sample by almost a factor of three. Using this unprecedented dataset, we interpret and model the vertical structure and kinematics across the LMC disc.} 
{We first split our parent sample into different stellar types (young and old). We then examine maps of vertical velocity, $v_{z'}$, moments (median and MAD) perpendicular to the LMC disc out to $R' \approx 5$ kpc; we also examine the vertical velocity profiles as a function of disc azimuth and radius. We interpret our results in the context of three possible scenarios: 1) time-variability in the orientation of the disc symmetry axis; 2) use of an incorrect LMC disc plane orientation; or 3) the presence of warps or twists in the LMC disc. We also present a new inversion method to construct a continuous 3D representation of the disc from spatially-resolved measurements of its viewing angles.} 
{Using young stellar populations, we identify a region in the LMC arm with highly negative $\overline{v_{z'}}$; this overlaps spatially with the supershell LMC~4. When interpreting the maps of $\overline{v_{z'}}$, our results indicate that: 1) the LMC viewing angles may vary with time due to, e.g., precession or nutation of the spin axis, but this cannot explain most of the structure in $\overline{v_{z'}}$ maps;
2) when re-deriving the LMC disc plane by minimising the RMS vertical velocity $v_{z'}$ across the disc, the inclination and line-of-nodes position angle are $i = 24^\circ$ and $\Omega = 327^\circ$, respectively, with an $\sim 3^\circ$ systematic uncertainty associated with sample selection, contamination, and the position of the LMC center; 3) when modelling in concentric rings, we obtain different inclinations for the inner and outer disc regions, and when modelling in polar segments we obtain a quadrupolar variation as function of azimuth in outer the disc. We provide 3D representations of the implied LMC disc shape. These provide further evidence for  perturbations caused by interaction with the SMC.}
{The combination of \textsl{SDSS}-IV/V and \gaia data reveal that the LMC disc is not a flat plane in equilibrium, but that the central bar region is tilted relative to a warped outer disc. }



\maketitle

\section{Introduction}
\label{sec:introduction}

The LMC is the Milky Way's (MW) most massive satellite galaxy and a member of the Local Group. Due to its close proximity \citep[$\approx50$ kpc,][]{pietrzynski19}, it provides astronomers with a unique view into the complexities of extragalactic systems. The LMC represents the prototype Barred Magellanic Spiral, a type of galaxies with unusual structural characteristics \citep[e.g.][]{devacouleurs-freeman72}. It is a dwarf, bulgeless spiral galaxy, with an off-centred and lopsided stellar bar, many star forming regions, and a single prominent spiral arm \citep[e.g.][]{Elmegreen1980,Gallagher1984,Zaritsky2004,Yozin2014, luri20,rathore25}. It is also a gas-rich galaxy \citep[e.g.][]{luks-rohlfs92,kim98}, characterised by an inclined and warped disc \citep[e.g.][]{vdmcioni01,vandermarel01,vdm02,Olsen2002, nikolaev04,Choi2018,ripepi22}.

The luminosity of the LMC is one-tenth of that of the MW \citep[e.g.][]{sparke00}, and its stars are distributed in a flat disc tilted at an inclination of $i \sim 30^\circ$ with respect to the line-of-sight, though there remains a large uncertainty in the literature on what the inclination angle is \citep[e.g.][]{vdm09,Haschke&Grebel&Duffau2012,vandermarel&kallivayalil2014,luri20,ripepi22}. The total (dynamical) mass of the LMC out to large radii is currently estimated to be around $M_{\text{LMC}}\approx 1.8 \times 10^{11}$\msun \citep[e.g.][]{penarrubia16,erkal19}, an order of magnitude larger than the mass concentrated in the regions where most of its stars are observed \citep[e.g.][]{avner-king67,vdm02}. 

Due to its proximity, the LMC is a perfect target for many dwarf galaxy studies and focused photometric surveys, such as \textsl{VMC} \citep{cioni11}, \textsl{SMASH} \citep{Nidever2017} or \textsl{VISCACHA} \citep{maia19}. Moreover, it is also one of the focuses of current and upcoming spectroscopic surveys, such as the Sloan Digital Sky Survey-V \citep[\textsl{SDSS}-V,][Kollmeier, J.A., et al. 2025, in prep.]{kollmeier19,almeida23}, the 4-metre Multi-Object Spectrograph Telescope \citep[\textsl{4MOST},][]{dejong19}, and future Large Survey of Space and Time \citep[\textsl{LSST},][]{ivezic19} at the Vera C. Rubin Observatory, as well as the astrometric mission \gaia (\textsl{ESA}).  \citet{helmi18,luri20} showed the capabilities of \gaia to characterise the structure and kinematics of this nearby galaxy using proper motions from \gaia DR2 and eDR3, building on earlier studies \citep[e.g.][]{vandermarel&kallivayalil2014,vdm16} using proper motions from the Hubble Space Telescope and \gaia DR1, respectively. 

The first to examine the internal kinematics of the LMC using \gaia DR3 data was \citet{jimenez-arranz23a}. Using \gaia DR3 proper motions, these authors presented an updated version -- with respect to the work of \citet{luri20} -- of the LMC in-plane velocity maps and profiles, using more than 10 million stars. Moreover, for $20-30$ thousand of these, \gaia supplied line-of-sight velocities, which were used to present and explore the first off-plane (vertical) velocity maps of the LMC disc. This was the first time that a homogenous data set with 3D velocity information was examined in detail for a galaxy other than the MW.  

As a follow-up, the same group aimed to understand the origin and amplitude of the LMC's vertical structure through tailored $N$-body simulations of isolated and interacting LMC-like galaxies \citep[\textsl{KRATOS} suite,][]{jimenez-arranz24b}. The authors pointed out that the LMC-SMC pericenters correlate well with a sudden increase in disc thickness, and the strength of this change
correlates with the pericenter distance, the disc instability, and
the merger mass (i.e., the SMC). With the interaction, the scale height has a peak. Then, after the disc has been heated, the thickness slightly decreases. Finally, the LMC-like disc relaxes to a larger scale height than the original, before the interaction.

Due to the limited available samples of line-of-sight velocities, earlier vertical kinematic studies were mostly limited to the innermost regions of the LMC's disc.  Thanks to the wider coverage provided by the last two phases of the Sloan Digital Sky Survey (\textsl{SDSS}), namely \textsl{SDSS}-IV/V, we are now in a position to extend such analysis across the LMC's entire disc. Moreover, with this expanded dataset it is then possible to perform
a comprehensive new investigation of the vertical structure and kinematics of the LMC disc
plane. This is the subject of this contribution.

This paper is organised as follows: in Sect. \ref{sec:data}, we describe
the datasets and LMC samples used throughout this work. In Sect. \ref{sec:vertical_velocity_maps}, we show the detailed LMC vertical kinematic analysis, showing the velocity maps and profiles of the different LMC samples. In Sect. \ref{sec:interpretation}, we aim to provide an interpretation of the LMC vertical velocity maps. In Sect. \ref{sec:discussion}, we contextualize our results by comparing them with other works in the literature. Finally, in Sect. \ref{sec:conclusions}, we summarise the main conclusions of this work.

\section{Data}
\label{sec:data}

Our dataset is comprised of a cross-match between the \textsl{SDSS}-IV/V and \textsl{Gaia} surveys. The \textsl{SDSS}-IV/V data is supplied from two phases of the Sloan Digital Sky Survey (\textsl{SDSS}), the fourth \citep[\textsl{SDSS}-IV,][]{blanton17} and fifth \citep[\textsl{SDSS}-V,][Kollmeier, J.A., et al. 2025, in prep.]{kollmeier17} phases, respectively. \textsl{SDSS}-IV data is comprised of all the Apache Point Observatory Galactic Evolution Experiment (\textsl{APOGEE}) survey data, compiled into the publicly available seventeenth data release \citep[DR17, ][Holtzman et al., in preparation]{Abdurrouf22}. Conversely, the Milky Way Mapper (\textsl{MWM}) survey of \textsl{SDSS}-V data comes from both \textsl{APOGEE} \citep{majewski17,wilson19} and the Baryon Oscillation Spectroscopic Survey \citep[\textsl{BOSS},][]{schlegel09} telescopes, and form part of the first ever data release by \textsl{SDSS}-V, the eighteenth data release \citep[DR18,][]{almeida23}. Lastly, all \textsl{Gaia} data come from the astrometric space mission satellite, and specifically from the third data release \citep[DR3, ][]{gaiadr3summary}.

The \textsl{APOGEE} survey, part of the third  \citep[\textsl{SDSS}-III,][]{eisenstein11}, fourth  \citep[\textsl{SDSS}-IV,][]{blanton17}, and fifth  \citep[\textsl{SDSS}-V, as part of the MWM;][]{kollmeier17} phases of \textsl{SDSS}, is an infrared spectroscopic survey sampling all Galactic stellar populations, from the inner bulge, throughout the disc, and in the MW halo. \textsl{APOGEE} is a dual-hemisphere survey that makes use of the 2.5-metre Sloan Telescope at Apache Point Observatory \citep[APO, \textsl{APOGEE}-North;][]{gunn06} as well as the 2.5-metre Du Pont telescope at Las Campanas Observatory \citep[LCO, \textsl{APOGEE}-South;][]{bowen-vaughan73}, each mounted with the \textsl{APOGEE} spectrograph. \textsl{APOGEE} collects high-resolution ($R\sim22~500$), near-infrared spectra for over $>1$ million stars in the MW. Of particular importance for this work is the fact that \textsl{APOGEE} has focused on targeting thousands of stars in the LMC/SMC galaxies, providing exquisite spectroscopic information in the form of precise line-of-sight velocities, stellar parameters, and element abundance ratios for bright Red Giant Branch (RGB) and Asymptotic Giant Branch (AGB) stars \citep[e.g.][]{nidever20,povick23c,povick24}.

In more detail, the raw \textsl{APOGEE} multi-fiber spectra are processed with the data processing pipeline \citep{nidever15b} producing 1-D extracted and wavelength calibrated spectra with accurate line-of-sight velocities, with typical errors of $\sim$0.1 \kms for LMC RGB stars \citep{nidever15b}. Line-of-sight velocities are determined with \texttt{Doppler} \citep{nidever21}, which is specifically optimised to improve the derivation of line-of-sight velocities by forward-modelling all of the visit spectra simultaneously with a consistent spectral model for faint sources with many visits. 

The \textsl{BOSS} survey \citep{schlegel09} uses an optical low-resolution ($R\sim2~000$) spectroscopic instrument. For \textsl{SDSS}-V's \textsl{MWM}, one of the optical \textsl{BOSS} spectrographs \citep[][]{schlegel09,dawson13,smee13} was relocated to LCO from APO, making it now a double-hemisphere survey. Moreover, \textsl{BOSS}' spectrograph contains 500 fibers, and observes in the optical (visible) regime (360-1000 nm). Despite the initial purpose of the \textsl{BOSS} programme being designed for cosmology, it is now collecting observations of stars in the MW and LMC/SMC. In \textsl{SDSS}-V, \textsl{MWM} is using \textsl{BOSS} to observe a variety of targets in the MW especially targeting hotter stars. The Magellanic Genesis survey is targeting 100,000 evolved giant stars across the entire face going to $G \sim 17.5$. Radial velocities, stellar parameters and some chemical abundances will be determined for these stars.

We combine the spectroscopic data of the \textsl{SDSS} survey with the astrometric (and spectroscopic) data from the \gaia DR3 \citep{gaiadr3summary}. The \gaia mission is a primarily astrometric (but now also with spectroscopic instruments) survey with a main goal to create the most precise and detailed 3D map of our Galaxy. Insofar, it has catalogued and determined astrometric and photometric data for almost two billion stars \citep{gaiadr2mission,gaiadr2summary,gaiaedr3summary,gaiadr3summary}, representing around $1\%$ of all stars of the MW. From the nearly two billion sources that \gaia has observed, around 15 million stars belong to the Clouds \citep{jimenez-arranz23a,jimenez-arranz23b}.

All together, this work aims to synergise the data from \textsl{SDSS} and \textsl{Gaia} surveys to map the vertical structure of the LMC. In the next few subsections, we describe the base \textsl{SDSS} (\textsl{APOGEE} and \textsl{BOSS}, Sect. \ref{subsec:SDSS_sample}) and \gaia samples (Sect. \ref{subsec:gaia_sample}), and the selection criteria that were used to create the LMC clean samples (also  Sect. \ref{subsec:gaia_sample}). The cross-match between the \textsl{SDSS} and the \gaia datasets that are used in this work is described in Sect. \ref{subsec:crossmatch}. Finally, in Sect. \ref{subsec:evolutionary_phases} we dissect different stellar populations of our LMC clean samples.

\subsection{SDSS sample}
\label{subsec:SDSS_sample}

We choose stars in the LMC sky area from the \textsl{SDSS}-V's \textsl{MWM} dataset (\textsl{BOSS} and \textsl{APOGEE}) and the \textsl{SDSS}-IV survey (\textsl{APOGEE} DR17) that have line-of-sight velocity information. We do so by restricting our selection to a region of 15\degr$~$radius around the LMC's photometric centre, defined as $(\alpha_c,\delta_c)$ = (81.28\degrnospace, --69.78\degrnospace) by \citet{vandermarel01}.

With this initial cut we obtain 55 845, 22 374, 11 328 stars for the \textsl{BOSS} MWM, \textsl{APOGEE} \textsl{MWM}, and \textsl{APOGEE} DR17 surveys, respectively. The \textsl{APOGEE} DR17 and \textsl{MWM} datasets will be combined into a single sample called \textsl{APOGEE} with a total of 33 702 stars. We shall also refer to the \textsl{BOSS} \textsl{MWM} sample as \textsl{BOSS} for simplicity.

\begin{figure*}[t!]
    \centering
    \includegraphics[width=1\textwidth]{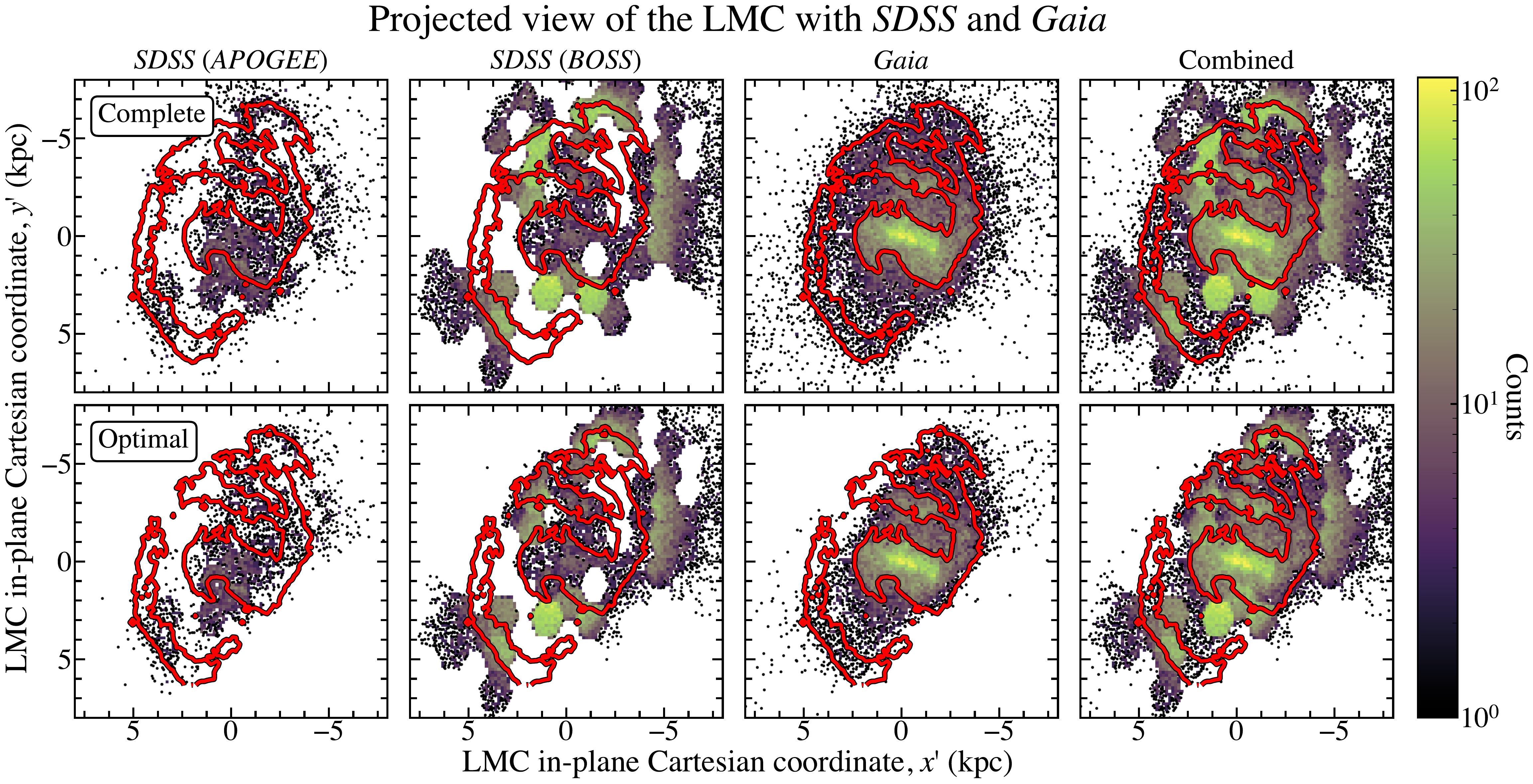}
    \caption{Comparison of the density maps between the different LMC clean samples. Top: LMC complete sample. Bottom: LMC optimal sample. From left to right: \textsl{APOGEE}, \textsl{BOSS}, \gaianospace, and Combined sample. We display bins containing three or more stars; otherwise, we display individual stars as a scatter plot. The brighter colours correspond to the higher density zones. A red line splitting the overdensities (LMC bar and spiral arm) from the underdensities is plotted. All maps are shown in the LMC in-plane $(x', y')$ Cartesian coordinate system. To mimic how the LMC is seen in the sky, the plotted data has both axes inverted.}
    \label{fig:samples_density}
\end{figure*}

Because of the selection function of the \textsl{BOSS} and \textsl{APOGEE} surveys, stars in these samples are not uniformly distributed in the sky, which makes the footprint -- composed of hexagonal patches -- highly visible (see, for example, Fig. \ref{fig:samples_density}). The reader may notice that this \textsl{SDSS} sample, with a total of 89 547 (\textsl{APOGEE} + \textsl{BOSS}) stars, contains both LMC and MW stars. The following Section (Sect. \ref{subsec:gaia_sample}) discusses how we obtain LMC clean samples, after removing the MW foreground contamination. We also note that, since we have enough coverage of the LMC's disc (Fig~\ref{fig:samples_density}), our results are unaffected by the uneven footprint imposed by the selection function.

\subsection{\gaia LMC clean samples}
\label{subsec:gaia_sample}

To remove foreground MW star contamination from our \gaia LMC base sample, we follow the neural-network (NN) technique from \cite{jimenez-arranz23a} and \cite{jimenez-arranz23b}, as a parallax based distance distinction has been shown to not be a reliable method due to their high uncertainties \citep[e.g.][]{luri20,Lindegren2021b}. This methodology enables the classification of three different samples of candidate LMC stars with different degrees of completeness and purity. Since in this work we aim to verify the robustness of our results with this trade-off between completeness and purity, we employ both the NN complete and NN optimal LMC clean samples from \cite{jimenez-arranz23a}. In a similar fashion to the \textsl{SDSS} samples, the \gaia LMC clean samples are restricted to a region of 15\degr~radius around the LMC centre, defined as $(\alpha_c,\delta_c)$ = (81.28\degrnospace, --69.78\degrnospace) by \citet{vandermarel01}.

The NN complete sample corresponds to the sample that prioritises not missing LMC stars at the price of a possible increased MW contamination. It contains 12~116~762 stars. Conversely, the NN optimal sample differs in the trade-off between completeness and purity, with less MW contamination than the NN complete sample but losing some LMC stars. It contains 9~810~031 stars. Both samples are dominated by older stellar populations \citep[see, e.g. Fig. 3 from][]{luri20}. We stress that, since the purity-completeness trade-off is a decision that will define the properties of the resulting sample and therefore can have an effect on the results obtained from it, we keep both samples for this study.

A sub-sample of bright stars \citep[$G \lesssim 17.65$,][]{deangeli23} of the LMC NN complete and NN optimal samples have \gaia DR3 line-of-sight velocity information. This is obtained with the Radial Velocity Spectrometer \citep[RVS, e.g.][]{katz04,cropper18,katz23}. The instrument is a near-infrared (845–872 nm), medium-resolution (R $\sim$ 11 500), integral-field spectrograph dispersing all the light entering the field of view. We refer to these sub-samples with \gaia RVS line-of-sight velocities as the corresponding \vlos sub-samples. The LMC complete and optimal \vlos sub-samples contain  30 749 and 22 686 stars, respectively.

As illustrated in Fig. 9 of \cite{jimenez-arranz23a} and in Fig. \ref{fig:LMC_vlos} of this work, if we assume that the MW stars have line-of-sight velocities $< 125$ \kmsnospace, then the MW contamination for the complete (optimal) \vlos sub-samples is approximately $\sim$5\% ($\sim$0\%). Since this subset only contains stars at the bright end ($G \lesssim 16$) of the \gaia sample, we can observe that it maximises the performance of the classifier.

\subsection{Cross-matching the \textsl{SDSS} data with the \gaia LMC clean samples}
\label{subsec:crossmatch}

We cross-match the \textsl{SDSS} sample obtained in this work with the LMC/MW classifier created by \cite{jimenez-arranz23a} in order to remove any MW contaminants. Through this cross-match between surveys, we are also able to provide to the \textsl{SDSS} sample the necessary astrometric and photometric data for this study. As previously mentioned, we retain both the LMC NN complete and NN optimal samples for this study because the purity-completeness trade-off determines the characteristics of the final sample and may therefore impact the result. Hereafter, we remove ``NN'' from the LMC complete and optimal sample for the purpose of simplicity.

The cross-match between the \textsl{SDSS} and the \gaia sample is done using \gaianospace's DR3 \sourceidnospace. Furthermore, every star in the \textsl{APOGEE} \textsl{MWM} sample has a match with the previous \gaia Data Release (\gaia DR2). In general, the only safe way to compare source records between different data releases is to check the records of proximal sources in the same small part of the sky, since it is not guaranteed that the same astronomical source will always have the same source identifier in different \gaia Data Release. The \texttt{gaiadr3.dr2\_neighbourhood} table of \gaia DR3 offers a precomputed crossmatch of these sources that accounts for the proper motions available in DR3. There may be none, one, or (rarely) many potential DR2 counterparts in the vicinity of a particular DR3 source. This occasional source confusion is an inevitable consequence of the merging, splitting and deletion of identifiers introduced in previous releases during the DR3 processing and results in no guaranteed one–to–one correspondence in source identifiers between the releases. In this work, if a \gaia DR3 star has multiple potential counterparts in DR2, we retain the closest match (based on sky position) for the cross-match between the \textsl{APOGEE} \textsl{MWM} and \gaia sample.

In order to account for the different \textsl{SDSS} and \gaia LMC clean samples, we define the following eight samples with full velocity information (both proper motion and line-of-sight velocities) for the remainder of the paper:

\begin{itemize}
    \item Sample 1 (2): \textsl{APOGEE} complete (optimal) sample;
    \item Sample 3 (4): \textsl{BOSS} complete (optimal) sample;
    \item Sample 5 (6): \gaia complete (optimal) sample;
    \item Sample 7 (8): Combined complete (optimal) sample.
\end{itemize}

Moreover, the stars in the Combined sample have line-of-sight velocities that are arranged to prioritise the source of the data in the following order; use the \textsl{APOGEE} line-of-sight velocity if an \textsl{APOGEE} observation is available. If not, consider the \textsl{BOSS} line-of-sight measurement. Finally, use the \gaia line-of-sight velocity if there is no \textsl{SDSS} (\textsl{APOGEE} or \textsl{BOSS}) line-of-sight measurement for this star. Since some stars have line-of-sight velocity measurements from multiple surveys, the number of stars in the Combined sample is less than the sum of the three samples (\textsl{APOGEE}, \textsl{BOSS}, and \gaianospace). 

\begin{figure}
    \centering
    \includegraphics[width=\columnwidth]{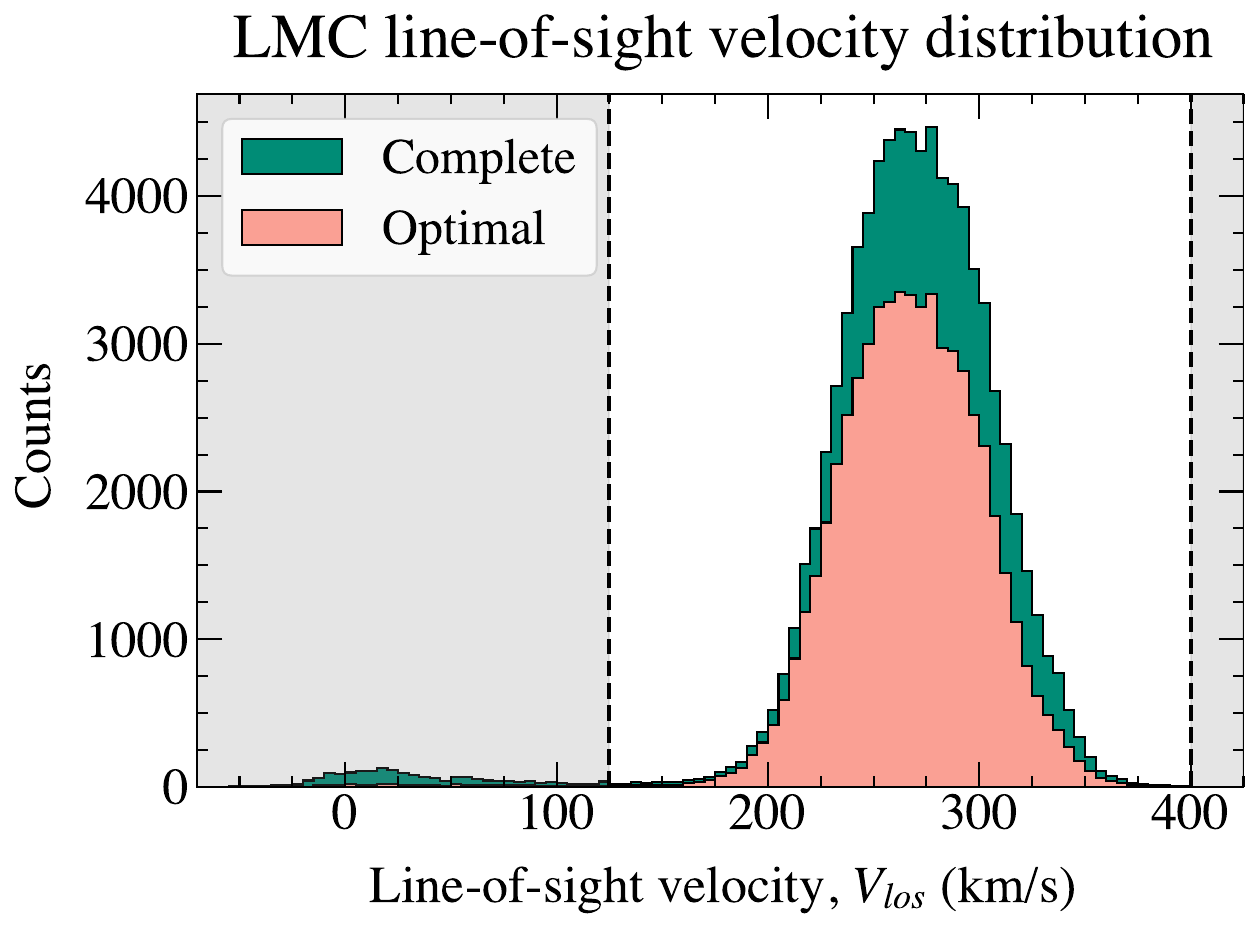}
    \caption{Distribution of the line-of-sight velocities of the LMC Combined complete (green) and optimal (salmon) sample. The vertical black dashed lines delimit the area of the LMC stars that are kept, namely, $125$ km/s $<V_{los}< 400$ km/s, to reduce MW foreground contamination.}
    \label{fig:LMC_vlos}
\end{figure}

Lastly, in order to reduce MW foreground contamination in our samples, we apply the following cut, $125$ km/s $<V_{los}< 400$ km/s, to select stars with line-of-sight velocities that are compatible with the LMC's systemic motion. The distribution of the line-of-sight velocities of the LMC Combined complete and optimal sample is displayed in Fig. \ref{fig:LMC_vlos}. Table \ref{tabl:samples} summarises the number of sources for the different LMC clean samples.

\begingroup
\setlength{\tabcolsep}{10pt} 
\renewcommand{\arraystretch}{1.5} 
\begin{table}
\centering
\begin{tabular}{lrr}
\hline \hline
LMC       &  Complete   & Optimal  \\ \hline
\textsl{APOGEE} \hspace{0.8cm}   &   6 162          & \hspace{1cm}   4 311      \\ 
\textsl{BOSS}      &   51 799          &    36 081      \\ 
\gaia     &   29 173        &    22 454      \\ 
Combined   &    80 453         &    58 509   \\ \hline
\end{tabular}
\caption{Comparison of the number of sources for the different LMC clean samples. Since some stars have line-of-sight velocity measurements from multiple surveys, the number of stars in the Combined sample is less than the sum of the three samples (\textsl{APOGEE}, \textsl{BOSS}, and \gaianospace).}
\label{tabl:samples}
\end{table}
\endgroup

\subsection{De-projection of coordinates into the LMC's disc plane}
\label{subsec:deproject}

Since the main goal of this work is to look at the vertical internal (off-plane) kinematics of the LMC, we use the coordinate transformation developed and used in \citet{vandermarel01, vdm02, jimenez-arranz23a} to de-project all coordinates into the LMC's disc reference frame. We assume that all stars lie in the $z'=0$ plane, and use the inclination, position angle, and position of the LMC centre given in \citet{jimenez-arranz23a}. Nonetheless, we re-derive the systemic motion for both proper motion and line-of-sight velocity by taking into account the median velocity value of a 0.5$^\circ$ inner circle in the LMC's disc, since asymmetries in the LMC are well-known \citep[e.g.][]{Olsen2002,Cullinane2022a,Cullinane2022b,jimenez-arranz23a,jimenez-arranz24a,kacharov24,rathore25}, as are the varying kinematics of different populations \citep[see, for example, Table 1 of][]{luri20}. Therefore, it is possible that the global rotation model fitted in \citet{luri20} to one sample of stars would yield a slightly different result from the actual velocities of stars in the vicinity of the adopted centre of a different sample, like the one used in this work -- see Section \ref{sec:vertical_velocity_maps} for the updated values for the systemic motions. The LMC centre matches its photometric centre \citep{vandermarel01} and is the same as that used in \citet{helmi18,luri20}.  Like any extra-galactic study of the kinematics of discs, we must make use of the infinitely thin disc approximation. This is because, unlike stars in the MW \citep{drimmel22}, or a few variable stars in the LMC \citep{cusano21,ripepi22}, the 3D position space of stars in the LMC is not available.

Figure \ref{fig:samples_density} shows the in-plane density distributions for the different LMC clean samples. To visualise the de-projected positions and velocities in the LMC in-plane $(x', y')$ Cartesian coordinate system for each star, we use the coordinate transformation described in \citet{vandermarel01}, \citet{vdm02} and \cite{jimenez-arranz23a}. In the top (bottom) panels, we show the LMC complete (optimal) \textsl{APOGEE}, \textsl{BOSS}, \gaianospace, and Combined sample, from left to right. A general characteristic across samples is that the LMC optimal sample does not cover the outer regions as much as the LMC complete sample. This trend is to be expected since the more complete we are, the more MW contamination we anticipate, and the classifier might experience issues in the LMC's periphery. A red line splitting the overdensities (LMC bar and spiral arm) from the underdensities is plotted for reference -- for more details see Sect 4.2 of \citet{jimenez-arranz23a}.

Stars in the inner region of the LMC, near the galactic bar, are observed in the \textsl{APOGEE} samples, while stars in the arm and the inter-arm region are observed by \textsl{BOSS}. As mentioned in Sect. \ref{subsec:SDSS_sample}, both samples show the footprint of the surveys' selection function, being more patchy for the \textsl{BOSS} sample (showing hexagonal shapes)\footnote{This is because the survey is not yet complete. As more data is collected, the missing patches observed across the LMC's disc will be filled in.}. One may observe that certain hexagonal patches, such as those centred in $(x',y')$ = (1, 3) kpc and (1, -5) kpc, exhibit an exceptionally high density in the LMC disc region throughout the complete \textsl{BOSS} sample. This is due to the fact that some LMC fields have high priority Transiting Exoplanet Survey Satellite \citep[\textsl{TESS}, ][]{ricker15} targets that need more epochs. Some of these high density patches in the \textsl{BOSS} samples disappear in the LMC optimal sample, as (x',y') = (1, -5) kpc, showing that they were likely mostly remnants of the MW contamination, whereas others, e.g.~at (x',y') = (1, 3) kpc, remain. The \gaia samples are relatively uniform, showing a decrease in density with radius, with a densely populated bar region and a sparsely populated arm area. After merging all these three samples (\textsl{APOGEE}, \textsl{BOSS} and \gaianospace), we obtain the LMC Combined samples, with a fair coverage of the whole LMC disc. 

Finally, we advise the reader that for most figures and analyses in this paper the two samples (LMC complete and optimal) yield qualitatively similar results, and we therefore generally show results only for the optimal sample. However, in cases where the two samples yield answers for a quantity of interest that differs by more than the statistical uncertainties, we discuss both results using the distinct LMC samples. The difference between the results for the two samples can then be used to assess any systematic uncertainties.

\subsection{Selecting stars at different evolutionary phases}
\label{subsec:evolutionary_phases}

Based on the unique properties of the various spectroscopic instruments, different evolutionary phases will be observed by each survey. To perform the LMC in-plane kinematic analysis for different stellar populations, we choose the target evolutionary phases by defining cut-outs with polygonal shape in the colour-magnitude diagram (CMD), as in \citet{luri20} -- see their Sect. 2.3 and Fig. 2. Even though it is not corrected from reddening, this rather crude selection can be used as an age-selected proxy to some extent. We define the stellar evolution areas as exclusive. Thus, they do not overlap. Based on the definition in \citet{luri20}, we consider the following evolutionary phases: 
\begin{itemize}
    \item Young 1: very young main sequence (ages < 50 Myr);
    \item Young 2: young main sequence (50 < age < 400 Myr);
    \item Young 3: intermediate-age main-sequence population (mixed ages reaching up 1-2 Gyr);
    \item RGB: red giant branch; 
    \item AGB: asymptotic giant branch (including long-period variables);
    \item RRL: RR-Lyrae region of the diagram;
    \item BL: blue loop (including classical Cepheids);
    \item RC: red clump.
\end{itemize}
 
\begin{figure*}[t!]
    \centering
    \includegraphics[width=0.65\textwidth]{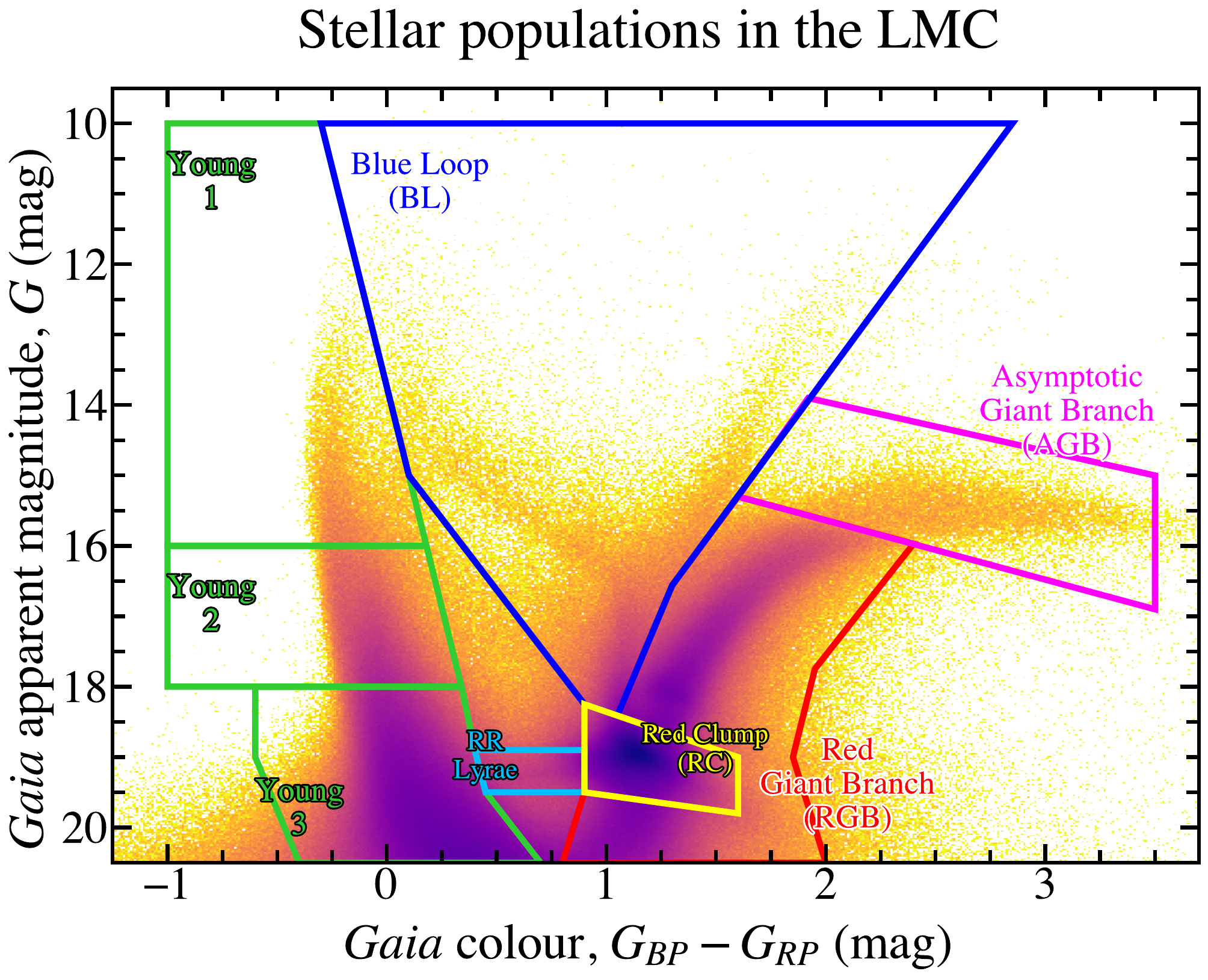}
    \includegraphics[width=1\textwidth]{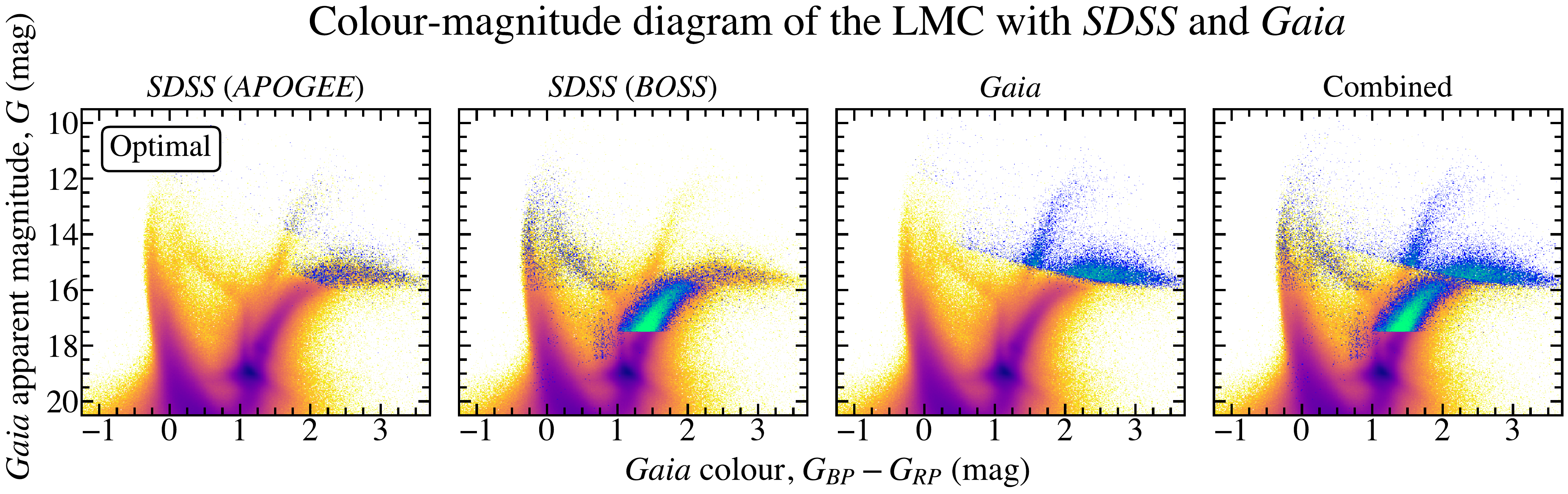}
    \caption{Top panel: colour-magnitude diagram (CMD) of the LMC optimal sample (9 810 031 stars) with the areas of the different evolutionary phases \citep[as defined by the polygons given in][]{luri20}. Second row: comparison of the CMD between the different surveys for the LMC optimal sample. From left to right: \textsl{APOGEE}, \textsl{BOSS}, \gaianospace, and Combined sample. In each sample, the foreground image represents the relative density of stars with line-of-sight velocity; higher densities are highlighted by lighter (greener) colours, while lower densities are shown by darker (bluer) colours. The background of the bottom panel displays the CMD for the optimal LMC sample, which consists of 9 810 031 stars. Colour represents the relative stellar density, with darker colours meaning higher densities.}
    \label{fig:samples_cmd}
\end{figure*}

The top panel of Fig. \ref{fig:samples_cmd} shows the \textsl{Gaia} CMD of the LMC optimal sample with the areas of the different evolutionary phases. The second row shows the CMD of the different surveys for the LMC optimal sample. We display, from left to right, the LMC optimal \textsl{APOGEE}, \textsl{BOSS}, \gaianospace, and Combined sample. These panels show that the majority of the stars that \textsl{APOGEE} observes are in the very bright evolutionary phases, like AGB and BL stars. While some AGB, BL, and Young stars are observed, the bulk of RGB stars are found in the \textsl{BOSS} survey. It is noteworthy that while the majority of LMC targets in \textsl{SDSS}-IV were RGB, \textsl{SDSS}-V is currently observing AGB stars. \gaia provides line-of-sight velocities for AGB and BL stars. Considering this, we can divide the Combined sample into distinct evolutionary phases that have ample coverage from \textsl{APOGEE}, \textsl{BOSS}, and \gaia data. We will consider the Young (hereafter, the joined Young 1, Young 2, and Young 3 samples), RGB, AGB, and BL Combined samples. The number of sources for each evolutionary phase of the LMC Combined sample and the contribution from each sample (\textsl{APOGEE}, \textsl{BOSS} and \gaianospace) is summarised in Table \ref{tabl:evol_phases}.

\begingroup
\setlength{\tabcolsep}{10pt} 
\renewcommand{\arraystretch}{1.5} 
\begin{table*}
\centering
\begin{tabular}{llrrrrr}
\hline \hline
 LMC      &  Evolutionary phase   &  \textsl{APOGEE}  &  \textsl{BOSS}  &  \gaia  &  \multicolumn{2}{c}{Combined}  \\ \cmidrule(rl){6-7} \hline
Complete   &   Young   &    3 \hspace{0.03cm}  &    1 404 \hspace{0.03cm}  &    0 \hspace{0.03cm}  &    1 356 \hspace{0.03cm}  &  1.7\%  \\ 
(Optimal)      &       &   (3)               &     (1 317)  &                  (0)  &                  (1 319)  &                (2.3\%)  \\ 
\hline
     &   BL        &    740 \hspace{0.03cm} &      4 313 \hspace{0.03cm}  &    6 285 \hspace{0.03cm} &    10 526 \hspace{0.03cm}  &  13.4\%  \\ 
     &             &   (530)  &                   (3 572)  &                  (5 150)  &                  (8 681)  &                (15.2\%)  \\
\hline
     &   AGB        &    4 705 \hspace{0.03cm} &    2 365 \hspace{0.03cm}  &    21 342 \hspace{0.03cm} &    23 092 \hspace{0.03cm}  & 29.5\% \\ 
     &             &    (3 247)  &                 (1 198)  &                  (16 141)  &                 (17 217)  &               (30.2\%) \\
\hline
     &   RGB        &    73 \hspace{0.03cm} &      43 274 \hspace{0.03cm}  &    42 \hspace{0.03cm}  &    43 374  \hspace{0.03cm} & 55.4\%   \\ 
     &             &    (55)  &                   (29 648)  &                  (31)  &                  (29 724)  &               (52.2\%) \\
\hline
\end{tabular}
\caption{Comparison of the number of sources for each evolutionary phase of the LMC Combined sample with sufficient statistics. We also show the contribution from each sample: \textsl{APOGEE}, \textsl{BOSS} and \gaianospace.}
\label{tabl:evol_phases}
\end{table*}
\endgroup

\section{Vertical velocity maps and profiles}
\label{sec:vertical_velocity_maps}

Thanks to the larger sample of line-of-sight velocities provided by \textsl{SDSS}-IV/V, we aim to provide a more extensive map of the LMC vertical (off-plane) velocity $v_{z'}$ than the one given in \cite{jimenez-arranz23a} using \gaia DR3 data -- see their Fig. 17. The coordinates transformations to deproject the LMC are summarized in Sect. \ref{subsec:deproject}. We advise the reader to consult \cite{vandermarel01}, \cite{vdm02} and \cite{jimenez-arranz23a} for additional information regarding the coordinate transformation and their underlying assumptions.

We use the inclination, position angle, and position of the LMC centre derived in \citet{luri20} and used in \citet{jimenez-arranz23a}. Nonetheless, we re-derive the systemic motion for both proper motion $(\mu_{\alpha*,0},\mu_{\delta,0})$ and line-of-sight velocity ($\mu_{z,0}$) for the LMC Combined complete and optimal sample by computing the median velocity value within a 0.5$^\circ$ with respect to the LMC center. The LMC Combined complete sample recovers a systemic motion of: $\mu_{\alpha*,0}$ = 1.914 $\pm$ 0.003 mas yr$^{-1}$; $\mu_{\delta,0}$ = 0.384 $\pm$ 0.002 mas yr$^{-1}$, and; $\mu_{z,0}$ = -1.113 $\pm$ 0.001 mas yr$^{-1}$. We adopt the same systemic motion for both LMC Combined complete and optimal since the variations are negligible -- differences lie within the uncertainties. Uncertainties are derived as MAD/$\sqrt{N}$, where MAD stands for median absolute deviation. Table \ref{tabl:systemic_motion} compares the systemic motions used in this paper with those used in \citet{luri20}.

\begingroup
\setlength{\tabcolsep}{10pt} 
\renewcommand{\arraystretch}{1.5} 
\begin{table}
\centering
\begin{tabular}{lll}
\hline
\hline
  &  Gaia Collab. (2021b)  & This work  \\
\hline
$\overline{\mu_{\alpha*}}$   & 1.858  & 1.914 $\pm$  0.002 \\
$\overline{\mu_{\delta}}$   & 0.385  & 0.384 $\pm$ 0.002  \\
$\overline{V_{los}}$   & 1.115   & 1.114 $\pm$ 0.003 \\
\hline
\hline
\end{tabular}
\caption{Comparison of the LMC systemic motion between this work and \citet{luri20}, expressed in mas yr$^{-1}$. Both systemic motions are referred to the LMC photometric centre $(\alpha_c,\delta_c)$ = (81.28\degrnospace, --69.78\degrnospace), by \citet{vandermarel01}. A kinematic model fit is used in \citet{luri20} to derive the systemic motion (see their Table 5, Main model). In this work, the systemic motion is re-derived by computing the median velocity within a 0.5$^\circ$ with respect to the LMC center, where uncertainties are derived as MAD/$\sqrt{N}$.}
\label{tabl:systemic_motion}
\end{table}
\endgroup

\subsection{LMC clean samples}
\label{subsec:vertical_velocity_maps_lmc_clean_samples}

\begin{figure*}[t!]
    \centering
    \includegraphics[width=1\textwidth]{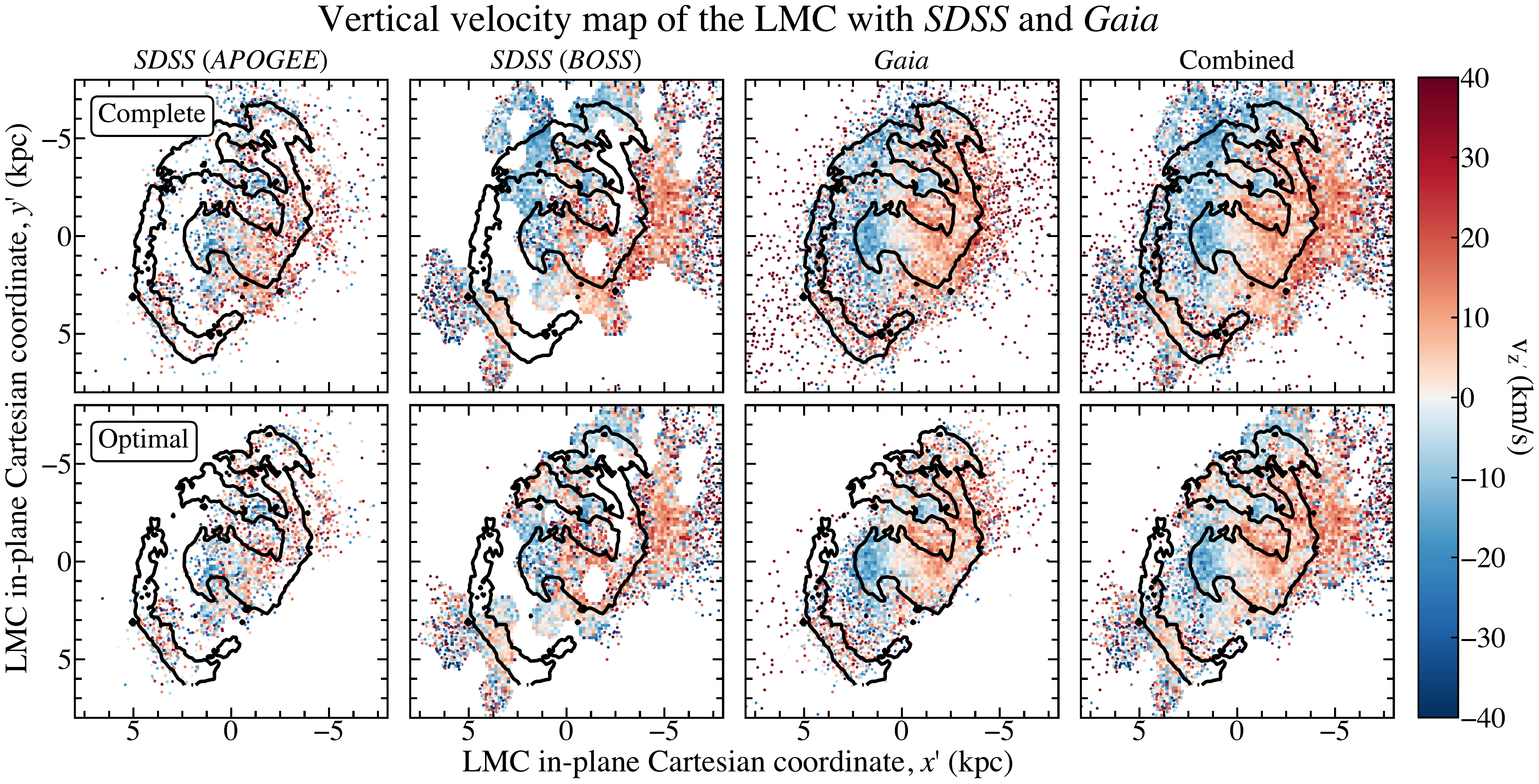}
    \caption{Comparison of the median vertical velocity maps between the different LMC clean samples. Top: LMC complete sample. Bottom: LMC optimal sample. From left to right: \textsl{APOGEE}, \textsl{BOSS}, \gaianospace, and Combined sample. We display bins containing three or more stars; otherwise, we display individual stars as a scatter plot. A black line splitting the overdensities (LMC bar and spiral arm) from the underdensities is plotted for reference. All maps are shown in the LMC in-plane $(x', y')$ Cartesian coordinate system. To mimic how the LMC is seen in the sky, the plotted data has both axes inverted.
    }
    \label{fig:samples_vertical_velocity}
\end{figure*}

Figure \ref{fig:samples_vertical_velocity} shows the median vertical velocity maps between the different LMC clean samples. The vertical velocity maps have second order differences between the LMC complete and optimal samples. The Combined samples exhibit a trend that is consistent with that reported by the \gaia sample \citep{jimenez-arranz23a}. A bimodal trend can be seen in the vertical velocity map, with half of the galaxy $(x' < 0)$ moving upward and the other half $(x' > 0)$ moving downward. Additionally, a gradient of increasing (median) vertical velocities from the inner to the outer disc appears to be positive (in absolute value). This could be associated with either: an overestimation of the disc inclination angle \citep[e.g.][]{jimenez-arranz23a}; the presence of a galactic warp \citep[e.g.][]{Choi2018,saroon22}; or due to the fact that the LMC is not in dynamical equilibrium \citep[e.g.][see Sect.~\ref{sec:interpretation}]{belokurov19,Choi2022,jimenez-arranz24a,jimenez-arranz24b}. Additionally, stars at the end of the bar with $x' > 0$ appear to be moving clearly in a negative direction, which may indicate that the bar is inclined with respect to the galactic plane \citep[e.g.][]{besla12,Choi2018}. 

\begin{figure}
    \centering
    \includegraphics[width=\columnwidth]{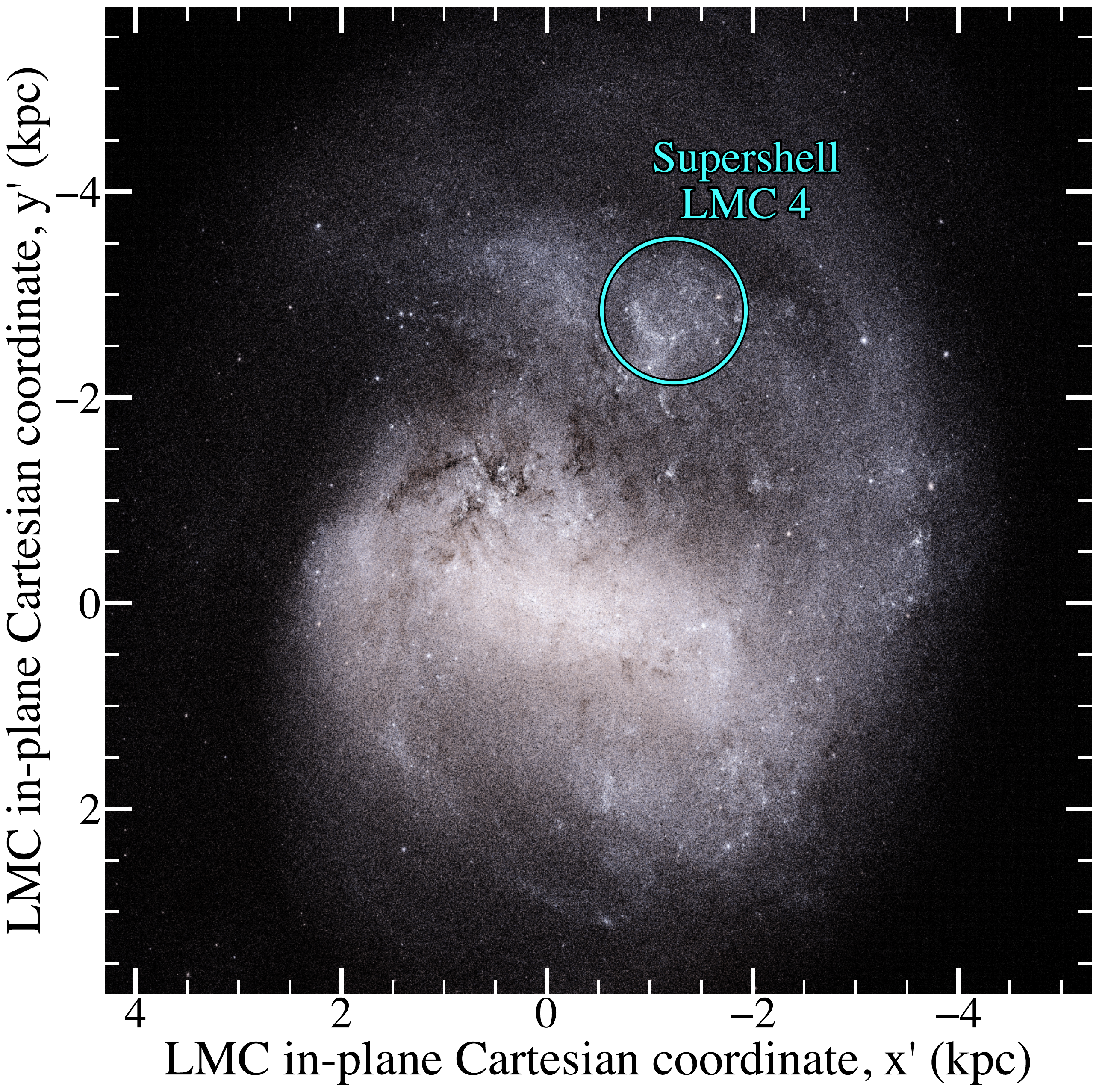}
    \caption{Representation of the LMC using astrometric and photometric information from the \gaia Data Release 3. This view is not a photograph since it has been compiled by mapping the total amount of radiation detected by \gaia in each pixel, combined with measurements of the radiation taken through different filters on the spacecraft to generate color information. The MW contamination has been removed because the NN Complete sample of \citet{jimenez-arranz23a} has been used. All types of stellar populations are shown. The cyan ring is centred on the supershell LMC~4 \citep[$\alpha = 5^{\text{h}}32^{\text{m}}$, $-66^\circ40'$,][]{book08}, with a diameter of 1,400 pc \citep{dopita85,ou24}. The representation is done in the LMC in-plane $(x', y')$ Cartesian coordinate system, as in Fig. \ref{fig:samples_vertical_velocity}. This represents a face-on view that is corrected for the viewing inclination.}
    \label{fig:supershell_LMC4}
\end{figure}

We detect a ``blob'' of stars with median negative vertical velocities\footnote{As shown in Fig. 11 of \citet{jimenez-arranz23a}, negative vertical velocities $v_{z'}$ represent a motion compatible with moving away from the observer (MW).} of around $\sim - 15$ \kms in the centre of the inner arm, at about $(x',y') = (-1, -3)$ kpc. This region corresponds to the supershell LMC~4 \citep[e.g.][]{vallenari03,book08,fujii14,ou24}, a structure of ionised material with size $\approx1$ kpc \citep[for a review read][]{tenorio-tagle88,dawson13b}; this is the largest bubble-like structure in the LMC. When projection effects are taken into account, the supershell LMC~4 is almost perfectly circular with a diameter of $\sim1400$ pc, and is expanding with a velocity of $\sim36$ \kms \citep[e.g.][]{dopita85,ou24}. Star formation appears to have started approximately $\sim15$ Myr ago.
We believe it is the first time that the supershell LMC~4 is observed in the (vertical) kinematics space of individual LMC stars. We will see later in Sect. \ref{subsec:vertical_velocity_maps_evolutionary_phases} how this kinematic feature is exclusive to the young stars. Figure \ref{fig:supershell_LMC4} is a representation of supershell LMC~4 -- in the LMC in-plane $(x', y')$ Cartesian coordinate system -- using astrometric and photometric data from the \gaia Data Release 3.

\subsection{Dependence on stellar evolutionary phases}
\label{subsec:vertical_velocity_maps_evolutionary_phases}

With the various evolutionary phases presented in Sect. \ref{subsec:evolutionary_phases}, we are able to construct the vertical velocity maps for the LMC Combined sample. In this way, the kinematic differences between different stellar tracers of the LMC can be analysed independently.

\begin{figure*}[t!]
    \centering
    \includegraphics[width=1\textwidth]{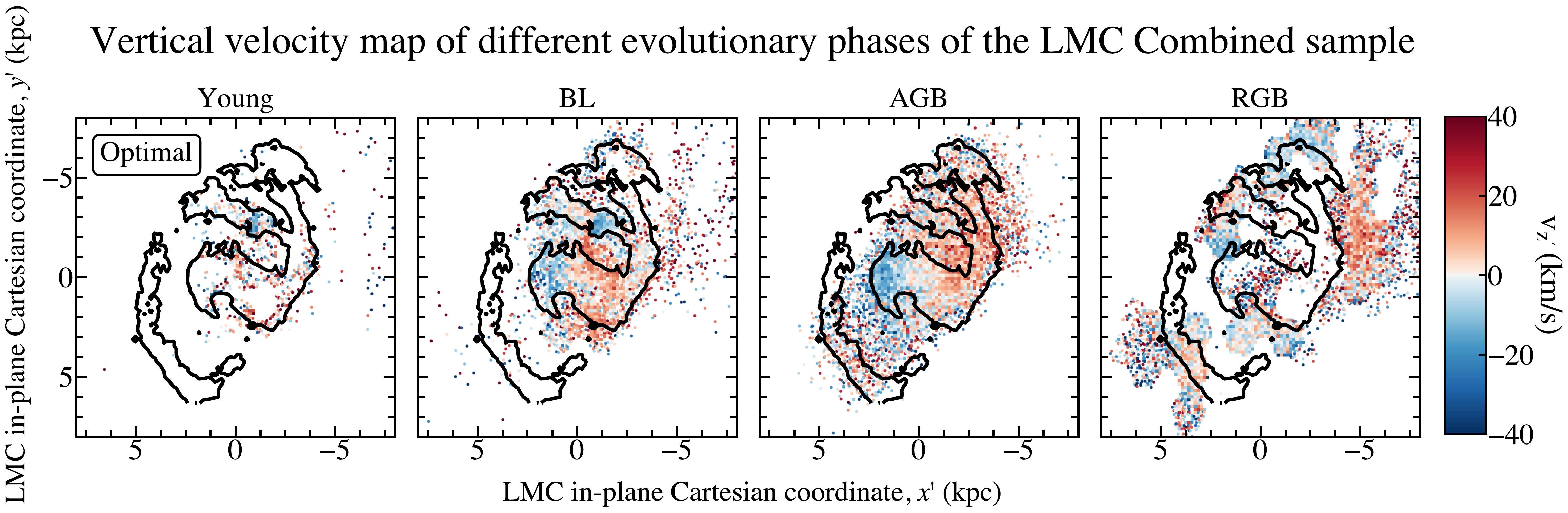}
    \caption{Comparison of the median vertical velocity maps between the different evolutionary phases (see Fig. \ref{fig:samples_cmd}) of the LMC Combined optimal sample. From left to right: Young, BL, AGB, and RGB samples -- following a rough sequence in age. We display bins containing three or more stars; otherwise, we display individual stars as a scatter plot. A black line splitting the overdensities (LMC bar and spiral arm) from the underdensities is plotted. All maps are shown in the LMC in-plane $(x', y')$ Cartesian coordinate system. To mimic how the LMC is seen in the sky, the plotted data has both axes inverted.}
    \label{fig:evol_vertical_velocity}
\end{figure*}

Figure \ref{fig:evol_vertical_velocity} shows the median vertical velocity maps between the different evolutionary phases of the LMC Combined optimal sample. In the top (bottom) panels, we show the LMC complete (optimal) Young, BL, AGB, and RGB samples, from left to right. Among all evolutionary phases, we observe that the Young sample is the least represented, due to target selection reasons. This population primarily follows the edges of the LMC bar, except for a few stars towards the east (i.e., the right side of Fig~\ref{fig:evol_vertical_velocity}), toward the position of the SMC. The supershell LMC~4 that was observed in Fig. \ref{fig:samples_vertical_velocity} and discussed in Sect. \ref{subsec:vertical_velocity_maps_lmc_clean_samples} is exactly traced by the Young population in the inner arm of the LMC. It shows its characteristic negative vertical velocity ($|v_{z'}|\sim-20$ km s$^{-1}$). The BL stars trace the LMC bar, the LMC's periphery, and the inner spiral arm, reiterating the kinematic imprint of the supershell LMC~4. For the BL stars, the area of the arm that is attached to the bar appears to be moving at a null vertical velocity, which is different from the positive vertical velocities seen in the older (AGB) stellar sample.

Including the bar and spiral arm(s), the AGB stars trace the entire LMC disc. The lack of the blue (negative vertical velocities) ``blob'' produced by the supershell LMC~4 is the main significant difference with the LMC Combined sample. Finally, the RGB stars, that are primarily observed by \textsl{BOSS}, appear unevenly distributed across the LMC disc due to the survey's footprint. This population also traces the inner part of the LMC bar. Due to the low number statistics for RGB stars in this region, we are unable to assess their vertical velocity profile.

\begin{figure*}[t!]
    \centering
    \includegraphics[width=1\textwidth]{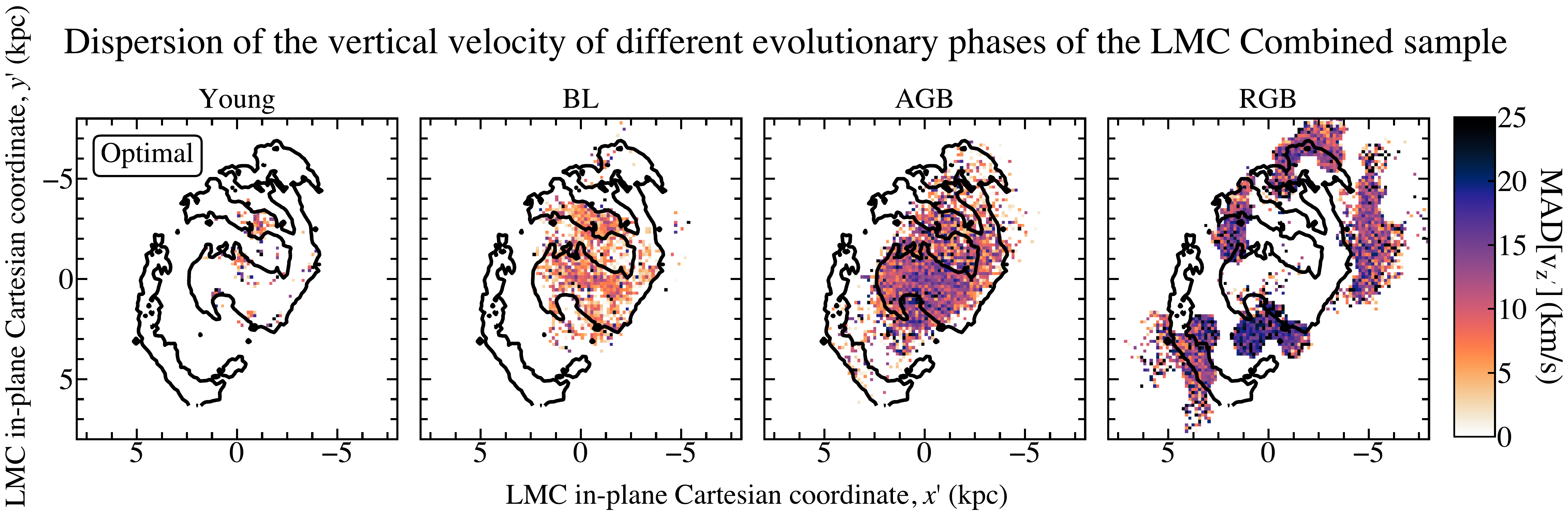}
    \caption{Same as Fig. \ref{fig:evol_vertical_velocity}, but for the vertical velocity dispersion -- given by the median absolute deviation (MAD). We only display bins containing three or more stars. Larger vertical velocity dispersion corresponds to darker colours.  }
    \label{fig:evol_vertical_velocity_dispersion}
\end{figure*}

In a similar vein, we use the median absolute deviation (MAD) to compute the dispersion of the vertical velocity of the different evolutionary phases of the LMC Combined sample, shown in Fig. \ref{fig:evol_vertical_velocity_dispersion}. The arrangement of the subplots is identical to that of Fig. \ref{fig:evol_vertical_velocity}. As expected, we find that the Young and BL samples, which comprise the younger evolutionary phases, exhibit a smaller velocity dispersion (of the order of $\sim$5 \kmsnospace) due to them being kinematically colder in comparison to the AGB and RGB samples, which comprise the older evolutionary phases. The AGB has a median vertical velocity dispersion around $\sim$12-15 \kmsnospace, whereas median dispersion of the RGB sample is above $\gtrsim$ 20\kmsnospace. In the region of the supershell LMC~4, the vertical velocity dispersion is smaller for the Young and BL sample than for the AGB sample. 

\subsection{Azimuthal analysis}
\label{subsec:vertical_velocity_maps_azimuthal analysis}

\begin{figure}[t!]
    \centering
    \includegraphics[width=1\columnwidth]{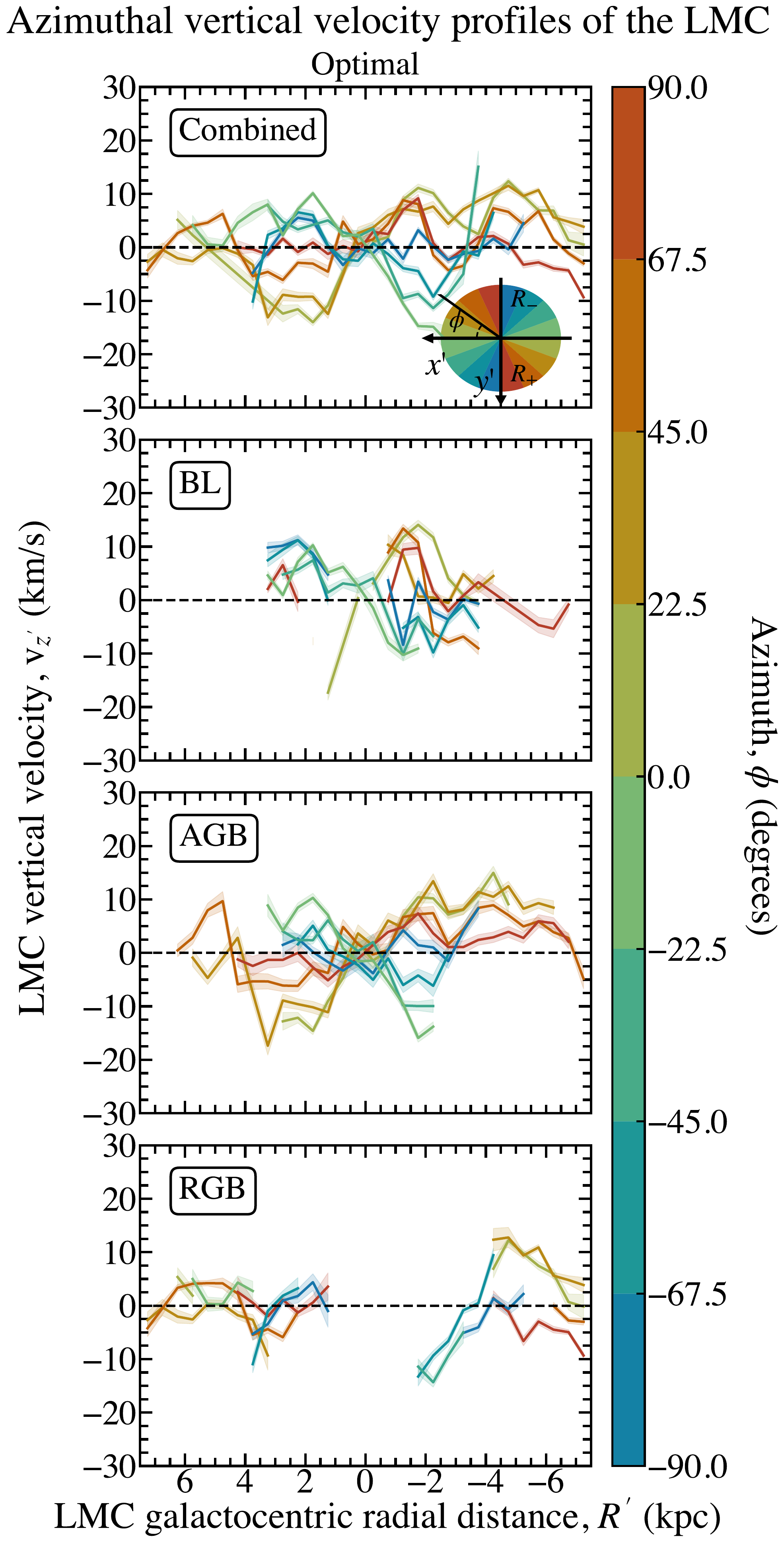}
    \caption{Comparison of the azimuthal vertical velocity profiles between the LMC optimal Combined sample and its the different evolutionary phases. From top to bottom: Combined, BL, AGB, and RGB samples. Because of its low statistical representation, the Young sample is not displayed. Only radial bins with more than 25 sources are plotted.  Each colour corresponds to a different azimuth $\phi$ defined in the LMC in-plane $(x', y')$ Cartesian coordinate system -- see schema in the top right panel. The vertical velocity dispersion of each bin, determined by the MAD, is displayed in the transparent shadows of each curve.}
    \label{fig:evol_1D_slices}
\end{figure}

In Fig. \ref{fig:evol_1D_slices} we present the one-dimensional vertical velocity profiles as a function of azimuth, $\phi$, defined in the LMC in-plane $(x', y')$ coordinate system. To do so, we bin the data in $16$ angular wedges, each of $22.5^{\circ}$ in width (see the schema in the top panel of Fig. \ref{fig:evol_1D_slices}). From top to bottom, we show the LMC optimal Combined, BL, AGB, and RGB samples. The Young sample is not shown due to low number statistics.

For the Combined sample we note that, near the LMC center ($R' \lesssim 1.75$ kpc), the azimuthal wedges along the line-of-nodes ($x' \sim 0$), namely, $\phi \sim 0 ^{\circ}$ and $\phi \sim 180 ^{\circ}$ with green lines, exhibit a distinct trend for the vertical velocity $v_{z'}$. The vertical velocity shows a linear increase (decrease) as function of the LMC galactocentric radial distance in the $\phi \sim 0 ^{\circ}$ ($\phi \sim 180 ^{\circ}$) wedge, as can be seen in the right panels of Fig. \ref{fig:samples_vertical_velocity}. On the other hand, the azimuthal wedges perpendicular to the line-of-nodes ($y' \sim 0$), namely, $\phi \sim 90 ^{\circ}$ and $\phi \sim -90 ^{\circ}$ with red and blue lines, show a flatter behaviour for all radius. The curves corresponding to the various wedges flatten at larger radii ($R' \gtrsim 4$ kpc), with similar dispersion for both the LMC Combined complete and optimal samples.

If we look at the different stellar populations, we are unable to determine whether the RGB stars exhibit the observed linear trend of vertical velocity with respect to the galactocentric radius due to the limited coverage of RGB stars in the LMC center. Nonetheless, it exhibits flattened curves with azimuth, and follows the same trend in the LMC outskirts as the LMC Combined sample. The situation is reversed for the AGB and BL stars, where the inner LMC region has better coverage than the outer regions. This allows us to see the same linear relationship near the centre of the LMC. In the following Section we provide a more thorough analysis of this kinematical behaviour.

\section{Interpretation of the LMC velocity maps}
\label{sec:interpretation}

In the previous Section we have shown that there is a dipole in the vertical velocity plane of the LMC disc (Figs.~\ref{fig:samples_vertical_velocity} and~\ref{fig:evol_vertical_velocity}). This may be due to different factors. In Sect. \ref{subsec:precession}, we examine whether this vertical velocity signature may be due to  a time-variability in the
LMC disc’s viewing angles. In Sect. \ref{subsec:incorrect_plane}, we examine whether this kinematic feature could be the result of using an incorrect LMC disc plane, given by the inclination, $i$, and line-of-nodes position angle, $\Omega$. In Sect. \ref{subsec:warps}, we examine whether we could be observing the imprint of a warp within the LMC disc. Finally, in Sect. \ref{subsec:LMC_center}, we discuss how the choice of the LMC center affects the robustness of our findings.

\subsection{The rate of change of the LMC's disc viewing angles}
\label{subsec:precession}

A dipole in the LMC vertical velocity maps could be explained by time-variability in the LMC disc viewing angles, namely, i.e., a non-zero rate of change $di/dt$ for the inclination and $d\Omega/dt$ for the line-of-nodes position angle $d\Omega/dt$. This would result in a solid body rotation of the LMC disc. 

The LMC rate of inclination change, $di/dt$, was first fitted by \citet{vdm02} using line-of-sight measurements of 1,041 carbon stars (see their Eq. 41). With new line-of-sight observations for both gas and stellar populations, subsequent works have attempted the same endeavour in a similar fashion \citep[e.g.][]{olsen07,olsen11}. However, the line-of-sight velocity field does not provide any information on time variations in the position angle of the line of nodes, $d\Omega/dt$. This makes it impossible to compute the past or future variation of the LMC symmetry axis orientation in an inertial frame tied to the MW.

The method used in \citet{vdm02} to estimate $di/dt$ was tailored to a case without astrometric information, and assuming circular orbits to make up for the absence of full 3D velocity info. Now, thanks to the spectro-astrometric data provided by the \gaia and \textsl{SDSS}-IV/V missions (see Sect. \ref{sec:data} for details), we have access to the vertical velocities for the LMC disc \citep[][and this work]{jimenez-arranz23a}. Under the assumption that these robustly describe the true vertical velocities across the LMC disc plane, this allows us to estimate both the rate of change of the LMC's disc inclination $di/dt$ and line-of-nodes angle $d\Omega/dt$ using a more direct method. 

Equation (6) of \citet{vdmcioni01} gives the LMC off-plane coordinate $z’$ as a linear sum of the LMC orthographic coordinates $(x, y, z)$ multiplied by trigonometric terms involving the inclination $i$ and line-of-nodes position angle $\Omega$. Then, we can take the time derivative of this equation at fixed $(x,y,z)$. This yields the LMC vertical velocity $vz’$ as a linear sum of $(x, y, z)$ multiplied by linear combinations of $di/dt$ and $d\Omega/dt$ that contain trigonometric terms involving the LMC viewing angles. Equation (4) of \citet{vdmcioni01} paper gives each of the LMC orthographic coordinates $(x, y, z)$ as a linear sum of the LMC deprojected coordinates $(x', y', z')$ multiplied by trigonometric terms involving the LMC viewing angles. If we substitute these equations for $(x,y,z)$ in the result of above, that yields the LMC vertical velocity $vz’$ as a linear sum of $(x', y', z')$ multiplied by linear combinations of $di/dt$ and $d\Omega/dt$ that involve trigonometric terms involving the LMC viewing angles. In this linear sum, most of the terms cancel and drop out, leaving: 
\begin{equation}\label{eq:model}
    v_{z'} = -x' \frac{d\Omega}{dt} \sin i + y' \frac{di}{dt}.
\end{equation}

The LMC vertical velocity maps $v_{z'}$ from the previous Section (see Fig. \ref{fig:samples_vertical_velocity}) can be modelled as a disc with solid body rotation defined by $di/dt$ and $d\Omega/dt$, which we can be fitted using this equation. For the LMC Combined complete sample, we thus obtain the rate of change of the LMC's disc viewing angles to be:
\begin{equation}
\begin{split}
    di/dt & = \hspace{0.1cm} 1.11 \pm 0.02 \text{ km s}^{-1} \text{ kpc}^{-1} \\
          & = \hspace{0.1cm} 65 \pm 1~^\circ \text{ Gyr}^{-1}, \\
    d\Omega/dt & = \hspace{0.1cm} 3.93 \pm 0.06 \text{ km s}^{-1} \text{ kpc}^{-1} \\
          & = \hspace{0.1cm} 230 \pm 4~^\circ \text{ Gyr}^{-1}.
\end{split}
\end{equation}

For the LMC Combined optimal sample:
\begin{equation}
\begin{split}
    di/dt & = \hspace{0.1cm} 0.76 \pm 0.03 \text{ km s}^{-1} \text{ kpc}^{-1} \\
          & = \hspace{0.1cm} 45 \pm 2~^\circ \text{ Gyr}^{-1}, \\
   d\Omega/dt & = \hspace{0.1cm} 2.83 \pm 0.07
 \text{ km s}^{-1} \text{ kpc}^{-1} \\
          & = \hspace{0.1cm} 166 \pm 4~^\circ \text{ Gyr}^{-1}.
\end{split}
\end{equation}

\begin{figure*}[t!]
    \centering
    \includegraphics[width=0.75\textwidth]{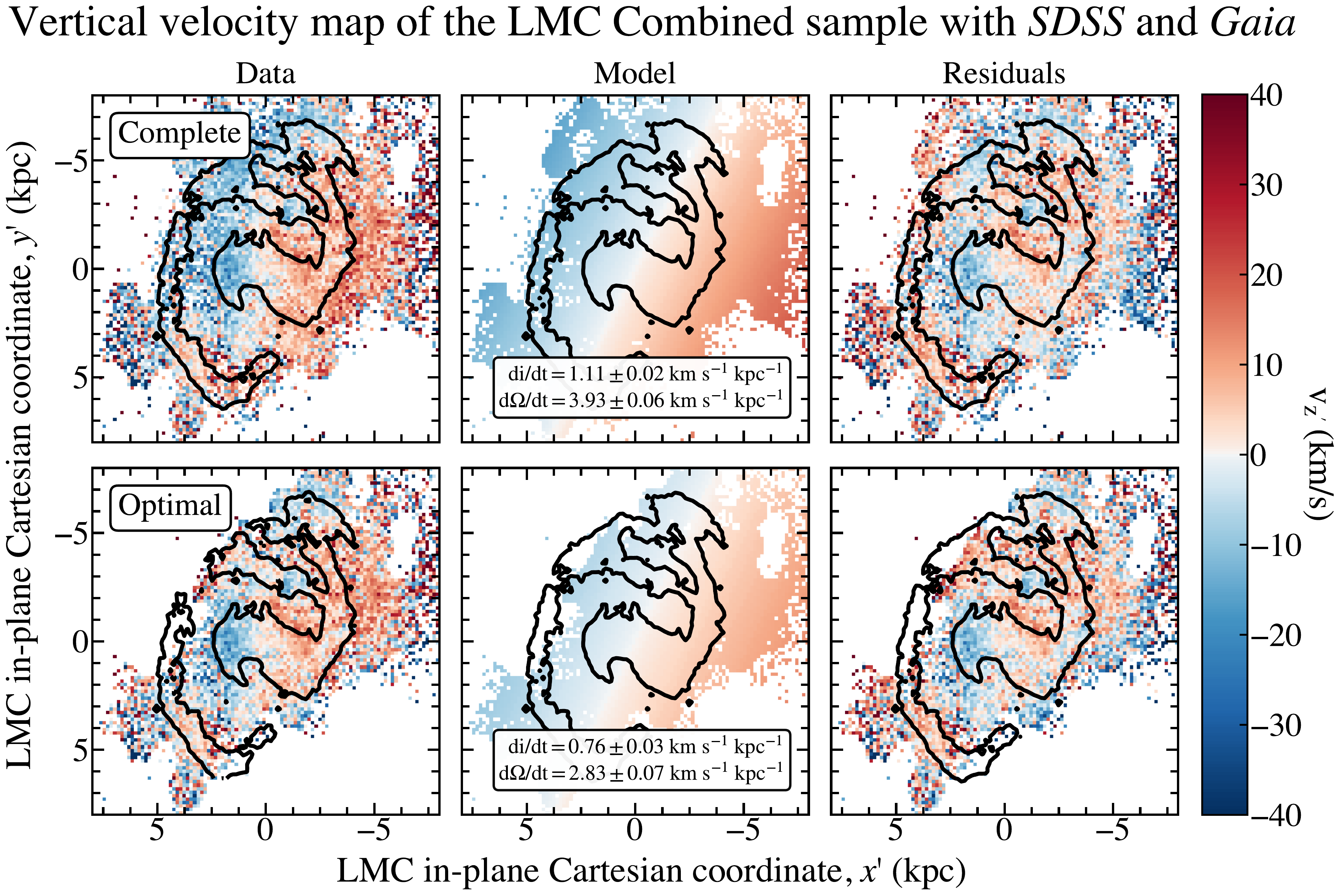}
    \caption{Comparison of the median vertical velocity maps between the different LMC Combined samples and the disc model with a solid body rotation. Top: LMC complete sample. Bottom: LMC optimal sample. From left to right: Data (Combined sample), model and residuals. We only display bins containing three or more stars. A black line splitting the overdensities (LMC bar and spiral arm) from the underdensities is plotted. All maps are shown in the LMC in-plane $(x', y')$ Cartesian coordinate system. To mimic how the LMC is seen in the sky, the plotted data has both axes inverted.}
    \label{fig:model_precession_nutation}
\end{figure*}

The statistical errors are computed by bootstrapping with replacement within the measured uncertainties of 1 000 pseudo-samples. Figure \ref{fig:model_precession_nutation} compares the median vertical velocity maps between the different LMC Combined samples, the disc model with a solid body rotation, and the residuals. Top (bottom) panels show the LMC Combined complete (optimal) sample. From left to right we show the vertical velocity maps for the data (same as right panels of Fig. \ref{fig:samples_vertical_velocity}), the disc model with a solid body rotation, and the residuals (data minus model). We can observe in the residuals maps that the inclusion of $di/dt$ and $d\Omega/dt$ helps to reduce the variance in the $vz’$ maps in the inner regions. However, the residual vertical velocity maps (right panels) clearly retain significant residual structure that cannot be attributed solely to non-zero values of $di/dt$ and $d\Omega/dt$. This implies that (a) other effects may be responsible for (or contributing to) the structure in the $v_{z’}$ maps (see Sects. \ref{subsec:incorrect_plane} and \ref{subsec:warps}); and/or (b) the random uncertainties by themselves are not a good measure of the true uncertainties in $di/dt$ and $d\Omega/dt$. 

The systematic uncertainties in our new determinations of $di/dt$ and $d\Omega/dt$ are at least as large as the difference between the results for the two samples. Our determinations of $di/dt$ and $d\Omega/dt$ are in fact entirely meaningless, if other effects are the main cause of the non-zero values and dipole in the $v_{z'}$ vertical velocity maps, which we explore in subsequent sections. We discuss the results obtained in this Section and compare our estimations of $di/dt$ and $d\Omega/dt$ with other values found in the literature in Sect. \ref{sec:discussion}.

\subsection{Assuming an incorrect disc plane}
\label{subsec:incorrect_plane}

Another possible explanation of the dipole present in the LMC vertical velocity maps could be that such measurements are non-physical. This feature may be produced by assuming an incorrect LMC disc plane, defined by the inclination, $i$, and line-of-nodes position angle, $\Omega$.

In the previous Sections, we assume the LMC disc plane fitted by \citet{luri20}, with $i=34^\circ$ and $\Omega=310^\circ$\footnote{Note that $\Omega$ differs from the angle $\Theta$ defined by \citet{vdmcioni01} and used by other authors by 180 degrees.}, as the true LMC disc plane. In this work, the authors fitted the LMC proper motion as a flat rotating disc with average $v_R = 0$ and $v_\phi = v_\phi (R)$, where $v_R$ and $v_\phi$ are galactocentric in-plane polar velocities, to determine the LMC disc plane; namely, assuming circular velocities. However, it is known that the LMC is intrinsically elongated \citep[e.g.][]{vandermarel01}. Therefore, the true LMC disc plane obtained under this assumption may be biased. Moreover, to add more complexity to this problem, in the literature we can find a range of estimates of the LMC inclination and line-of-nodes position angle provided by different methods \citep[e.g.][]{vandermarel&kallivayalil2014,Haschke&Grebel&Duffau2012,ripepi22,kacharov24}, with differences that can be as large as 10$^\circ$. Thus, these two crucial assumed quantities are by no means a well constrained measurement.

In this Section, to determine whether the bimodal trend seen in the LMC vertical velocity maps could be the result of an incorrect disc plane assumption, we fit instead the LMC disc plane  by minimizing  the RMS vertical velocity, $v_{z'}$, across the disc of the LMC. To do so, the statistic that we minimise to find the best fitting values of the inclination and line-of-nodes position angle is a chi-squared, that can be explicitly written as follows:
\begin{equation}
    \chi^2 = \sum_{j=1}^{N_{\text{stars}}} (v_{z',j})^2.
\end{equation}
We remind the reader that the detailed relation $v_{z'} = v_{z'}(i,\Omega)$ can be found in \cite{vandermarel01}, \cite{vdm02} and \cite{jimenez-arranz23a}. We use the \texttt{SciPy} \citep{scipy20} implementation of the Nelder-Mead simplex algorithm \citep{nelder-mead65} to minimise $\chi^2$.

For the LMC Combined optimal (complete) sample, we recover a disc plane with inclination $i = 23.6^\circ\pm0.1^\circ$ ($i=20.0^\circ\pm0.1^\circ$) and line-of-nodes position angle $\Omega = 326.9^\circ \pm 0.3^\circ$ ($\Omega = 330.0^\circ \pm 0.2^\circ$). The statistical errors are computed via bootstrapping with replacement of 25 pseudo-samples. Moreover, to ensure the robustness of the optimisation, each pseudo-sample run started with a different initial guess of $(i,\Omega)$. It is worth mentioning that both the LMC Combined complete and optimal samples provide a robust estimation of $(i,\Omega)$ with a difference of $\sim3^\circ$, which can be considered as the systematic error. Finally, we observe that, in comparison to the values obtained by the authors of \cite{luri20}, namely $i = 34^\circ$ and $ \Omega=310^\circ$,  the LMC disc plane inclination $i$ and line-of-nodes position angle $\Omega$ that we recover for the LMC Combined optimal (complete) sample differ by $\delta i \sim 10^\circ$ ($\delta i \sim 14^\circ$) and $\delta \Omega \sim 17^\circ$ ($\delta \Omega \sim 20^\circ$). When compared to the LMC viewing angles of \citet{kacharov24}, also based on 3D kinematics, their estimate of the inclination ($i = 25.5^\circ \pm 0.2 ^\circ$) is similar to ours, while their estimate for the line-of-nodes position angle ($\Omega = 304^\circ \pm 0.4 ^\circ$) is not. We address the potential reasons for these discrepancies in Sect. \ref{sec:discussion}.

\begin{figure}[t!]
    \centering
    \includegraphics[width=\columnwidth]{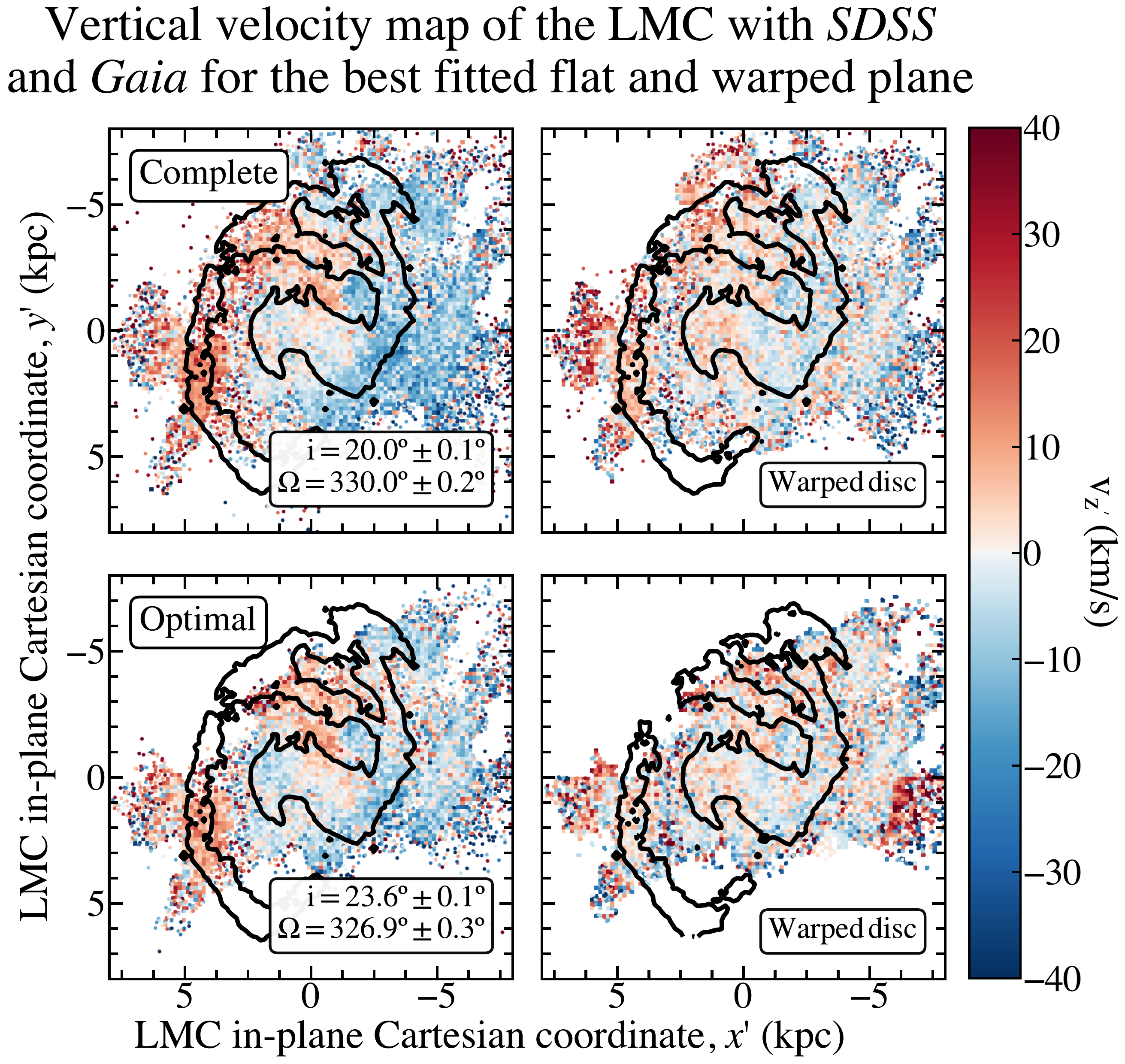}
    \caption{Comparison of the median vertical velocity maps for the different LMC Combined samples (top: complete sample; bottom: optimal sample), obtained with different assumptions about the LMC disc plane. Left: for the best-fitting flat plane, obtained by minimizing the RMS vertical velocity $v_{z'}$, with viewing angles listed in the panels. Right: for the best best-fitting warped plane, with viewing angles shown in Figure~\ref{fig:disc_fit_vs_azimuth}. When comparing the top and bottom panels, the reader should note that the overdensity lines differ slightly because each LMC sample uses a different disc plane.}
    \label{fig:maps_correct_plane}
\end{figure}

The left panels of Figure \ref{fig:maps_correct_plane} show the median vertical velocity maps for the LMC Combined complete (top) and optimal (bottom) samples, in a similar fashion to right panels of Fig. \ref{fig:samples_vertical_velocity}, but for the best LMC disc plane fitted by minimizing the RMS vertical velocity. When comparing the kinematic maps of the assumed \citep{luri20} and fitted values (this work) of $i$ and $\Omega$ (i.e., Fig. \ref{fig:samples_vertical_velocity} and Fig. \ref{fig:maps_correct_plane}), we observe the following features: $1)$ the LMC bar major axis becomes less aligned with the line-of-node due to the change of plane -- note there is no physical reason that the two should be aligned; $2)$, the LMC bar's bimodal vertical velocity trend vanishes, leaving behind a nearly null median vertical velocity; $3)$ there is a sign change in the disc's vertical velocity bimodal trend. Compared to the positive values in Fig. \ref{fig:samples_vertical_velocity}, the part of the spiral arm attached to the bar now shows a negative vertical velocity. Lastly, and in connection with the previous point, $4)$ the supershell LMC~4 continues to stand out from the background, albeit not as much as with the LMC disc plane fitted in \cite{luri20}. 

While the newly determined disc plane (by definition) provides a lower RMS vertical velocity than the one determined by \citet{luri20}, the median maps in 
the left panels of Figure \ref{fig:maps_correct_plane}
continue to show significant residual structure. This implies that the underlying assumption of this Section --that the structure observed in the vertical velocity maps is solely due to incorrect constant viewing angles-- is incomplete. Something else must be responsible for the majority of the vertical velocity structure, which we examine in the next Section.

\subsection{Warp in the LMC disc}
\label{subsec:warps}

Since the LMC was previously found to contain a warp \citep[e.g.][]{vdmcioni01,Olsen2002,nikolaev04,Choi2018,ripepi22}, it is important to assess the validity of the assumption in Sect. \ref{subsec:incorrect_plane} that the disc plane can be characterized by constant values of $(i,\Omega)$. 

\begin{figure}
    \centering
    \includegraphics[width=\columnwidth]{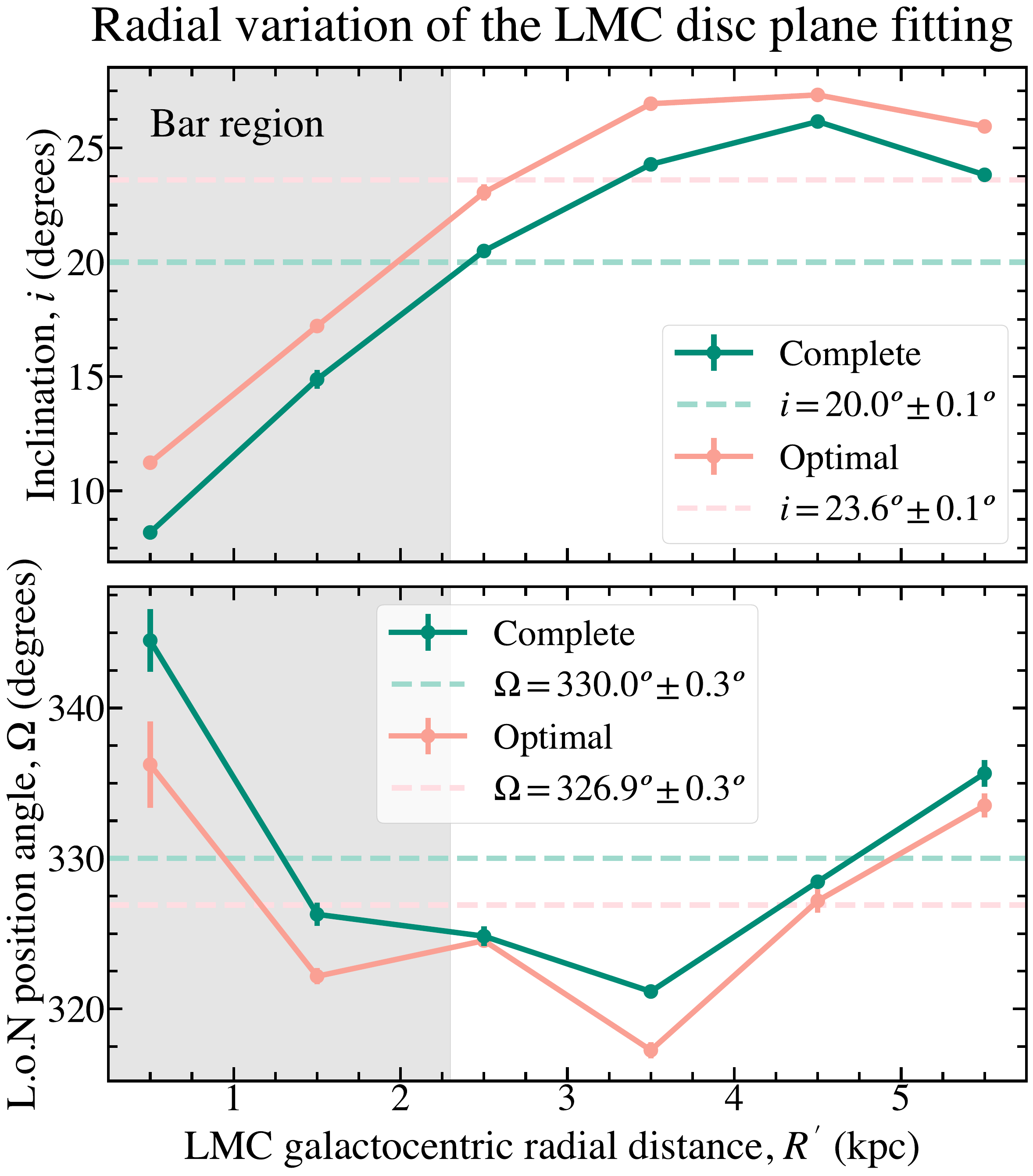}
    \caption{Variation of the LMC disc plane fitting as function of the LMC galactocentric radius $R'$. Top: Inclination $i$. Bottom: Line-of-nodes position angle $\Omega$. The ``global'' fit derived in Sect. \ref{subsec:incorrect_plane} is displayed by the horizontal dashed lines. We use a grey shaded area to draw attention to the bar region \citep[$R'_{\text{bar}}=2.3$ kpc,][]{jimenez-arranz23a}. In all panels we display the LMC complete (optimal) sample in green (salmon).}
    \label{fig:disc_fit_vs_radius}
\end{figure}

First, for distinct annular rings in the $(x',y')$ plane -- defined by the inclination angle, $i$, and line-of-nodes position angle, $\Omega$, of \cite{luri20}, Sect. \ref{sec:vertical_velocity_maps} --, we repeat the best fit disc plane minimisation of Sect. \ref{subsec:incorrect_plane} to quantify the change of $(i,\Omega)$ as function of the LMC galactocentric radial position $R'$. Figure \ref{fig:disc_fit_vs_radius} (left panels) displays the variation of the fitted LMC disc plane as function of the LMC galactocentric radius, $R'$. In the top (bottom) panel, we show the derived inclination $i$ (line-of-nodes position angle $\Omega$) for both the LMC Combined complete and optimal samples in green and salmon colours, respectively. The horizontal dashed lines show the ``global'' fit found for each sample in Sect. \ref{subsec:incorrect_plane}. In the top panel, we observe that in both samples the inclination is low ($i \sim 10^\circ$) at the inner regions ($R' \sim 0$ kpc) and grows up to values of $i \sim 25^\circ$ for the outer disc ($R' \gtrsim 3.5$ kpc). The LMC Combined complete and optimal samples have an almost constant off-set of approximately $\delta i \sim 2^\circ-3^\circ$, with the LMC Combined optimal sample having a systematically larger inclination. The bottom panel illustrates that the line-of-nodes position angle in both samples is high ($\Omega \sim 335^\circ - 350^\circ$) at inner regions ($R' \sim 0$ kpc), decreases to values of $\Omega \sim 320^\circ - 325^\circ$ at approximately $R' \sim 2.5$ kpc, and then increases once more to values of $\Omega \sim 335^\circ$ in the outer disc ($R' \gtrsim 3.5$ kpc). For nearly all LMC galactocentric radii, there is an almost constant off-set of approximately $\delta \Omega \sim 5^\circ$ between the LMC Combined complete and optimal sample, with the LMC Combined complete sample displaying a larger line-of-nodes. We highlight with a grey shaded area the LMC bar region, $R'_{\text{bar}} = 2.3$ kpc, found by \cite{jimenez-arranz23a}.

\begin{figure*}[t!]
    \centering
    \includegraphics[width=1\textwidth]{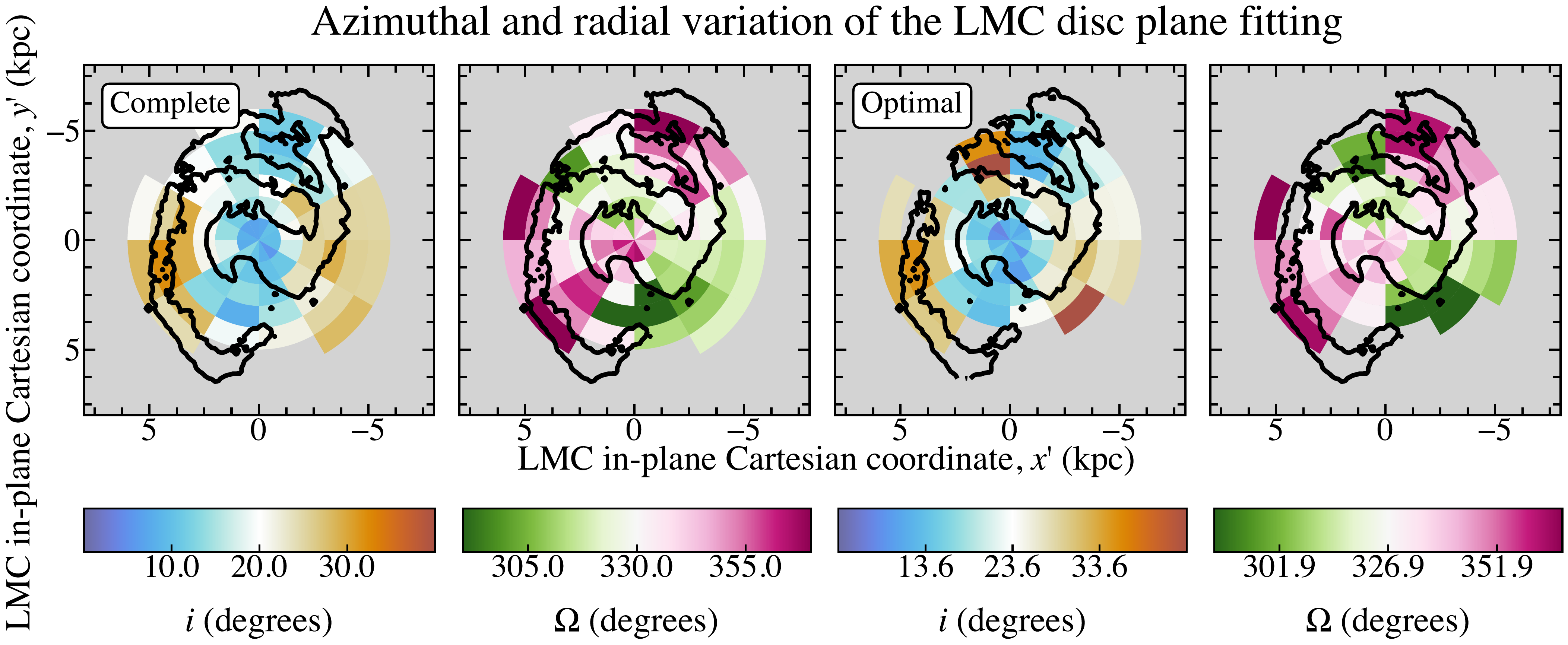}
    \caption{Variation of the LMC disc plane fitting as function of the azimuth $\phi$ and the LMC galactocentric radius $R'$. Left panels: LMC Combined complete sample. Right panels: LMC Combined optimal sample. The first and the third (second and fourth) panels show the inclination $i$ (line-of-nodes position angle $\Omega$). Only bins with more than 50 stars are displayed. Each colorbar is centred in the ``global'' fit derived in Sect. \ref{subsec:incorrect_plane}.}
    \label{fig:disc_fit_vs_azimuth}
\end{figure*}

Second, we can also quantify the change of the viewing angles $(i,\Omega)$ as a function of both the LMC azimuth, $\phi$, and galactocentric radius, $R'$, by repeating the best fit disc plane in annular wedges. In this manner, if there is an asymmetric warp in the LMC disc, it will be clearly manifested in this test. The results from this exercise are illustrated in Figure \ref{fig:disc_fit_vs_azimuth}. The left (right) panels are for the LMC Combined complete (optimal) sample, whereas the first and the third (second and fourth) panels show the best fit for the inclination $i$ (line-of-nodes position angle $\Omega$). First, we note that, in agreement with Fig. \ref{fig:disc_fit_vs_radius}, the LMC Combined complete and optimal sample both display a fairly homogeneous and isotropic central part of the disc with low inclination. However, we observe a quadrupole pattern in the outer region of the LMC disc (around $R'\gtrsim2$ kpc), with a significant variation as a function of the azimuth $\phi$. For the LMC optimal sample, the disc outskirts exhibit slightly higher inclinations. Second, both LMC samples exhibit a quadrupole pattern in the outer disc region, with a large line-of-nodes position angle in the inner regions.

The right panels of Figure~\ref{fig:maps_correct_plane} show the median vertical velocity maps for the LMC Combined complete (top) and optimal (bottom) samples for the best warped disc plane (see Fig.~\ref{fig:disc_fit_vs_azimuth}). The fitting of an LMC plane while allowing for both azimuthal and radial variation significantly decreases the structure in the maps. If we bin the maps onto a $0.5 \times 0.5$ kpc$^2$ grid, the RMS among the grid values drops by $\sim 50$\% (comparing the right panels of Figs.~\ref{fig:maps_correct_plane} and~\ref{fig:samples_vertical_velocity}). Instead, fitting for time-varying viewing angles (Sect.~\ref{subsec:precession}) or a new best-fit flat plane (Sect.~\ref{subsec:incorrect_plane}) only reduces this RMS by typically $\lesssim 20$\%. This suggests that a warp in the LMC disc is the most likely cause of the structure we see in the vertical velocity maps.

\subsection{3D models of the LMC disc}

\begin{figure*}[t!]
    \centering
    \includegraphics[width=1\textwidth]{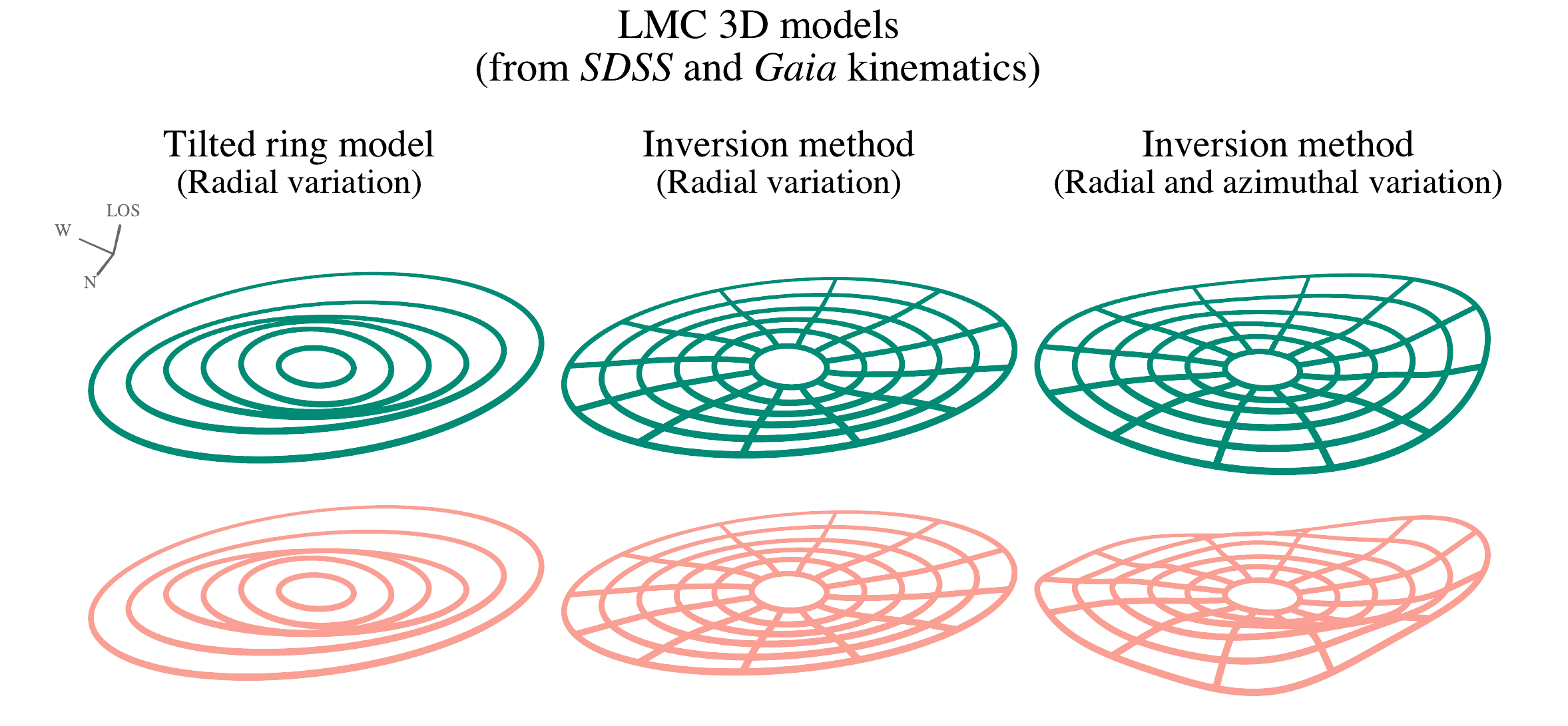}
    \caption{Comparison of 3D models for the LMC disc mid-plane based on the inferred spatial variations in viewing angle. Left panels: Tilted ring model. Central panels: Inversion method to best fit the inferred viewing angle variations with galactocentric radius $R'$ as a continuous surface. Right panels: Inversion method to best fit the inferred viewing angle variations with both galactocentric radius $R'$ and azimuth $\phi$ as a continuous surface. Top panels (green): LMC Combined complete sample. Bottom panels (salmon): LMC Combined optimal sample. We plot a triad that displays the unit vectors in the following directions: W, N, and LOS, being the latter the observer's direction. In the (x,y,z) frame, those vectors are defined by (1,0,0), (0,1,0), and (0,0,1), respectively -- see Fig. 11 of \citet{jimenez-arranz23a}.  A video of these LMC 3D models viewed from different directions is available online.}
    \label{fig:3d_model}
\end{figure*}

Now that we have determined the viewing angles as function of position, we wish to use this information to construct a three-dimensional (3D) model of the LMC disc. One traditional way to do this is to use the annular fits in Fig. \ref{fig:disc_fit_vs_radius}, and to represent the LMC disc as a collection of tilted rings. Models based on tilted rings have been used extensively to describe the structure of galaxies, especially by fitting the line-of-sight kinematics of the rings (their rotational and systemic velocities) to the velocity field observed in neutral hydrogen \citep[e.g.][]{rogstad74,corbelli10,boisvert16} or other tracers \citep[e.g.][]{abedi14,jones17}. In the present paper this kinematic fitting is supplanted by the 3D kinematics modelling that underlies the results in Fig. \ref{fig:disc_fit_vs_radius}. In the left panels of Fig. \ref{fig:3d_model}, we show the LMC tilted ring model using the derived inclination $i$ and line-of-nodes position angle $\Omega$ as function of galactocentric radius $R'$ from this Section -- see Fig. \ref{fig:disc_fit_vs_radius}. We display the LMC Combined complete and optimal samples with green and salmon rings, respectively\footnote{A video of the LMC 3D models from different viewing angles is made available online. Additionally, \url{https://www.oscarjimenezarranz.com/visualizations/interactive-lmc-3d-models} offers an interactive version of the LMC 3D models.}

Tilted ring models have several disadvantages. First, the disc is not represented as a continuous surface, which is unphysical. Second, it is not self-evident how to represent the disc in 3D when the viewing angles are known for disjunct polar segments as in Fig~\ref{fig:disc_fit_vs_azimuth}, instead of closed annuli. We therefore developed a new inversion method that takes on input the determinations of the viewing angles on a grid of polar segments (or annuli), and  returns on output the continuous surface $z'(x',y')$ that best reproduces these viewing angles in a least-squares sense. This method is described in the Appendix. The middle panels of Fig.~\ref{fig:3d_model}, show the best-fit LMC disc surfaces obtained with this method given the annular viewing angle results in Fig.~\ref{fig:disc_fit_vs_radius}; similarly, the right panels show the surfaces obtained given the polar-segment viewing angle results in Fig~\ref{fig:disc_fit_vs_azimuth}. 

\begin{figure*}
    \centering
    \includegraphics[width=0.7\textwidth,trim={6cm 0cm 6cm 0cm},clip]{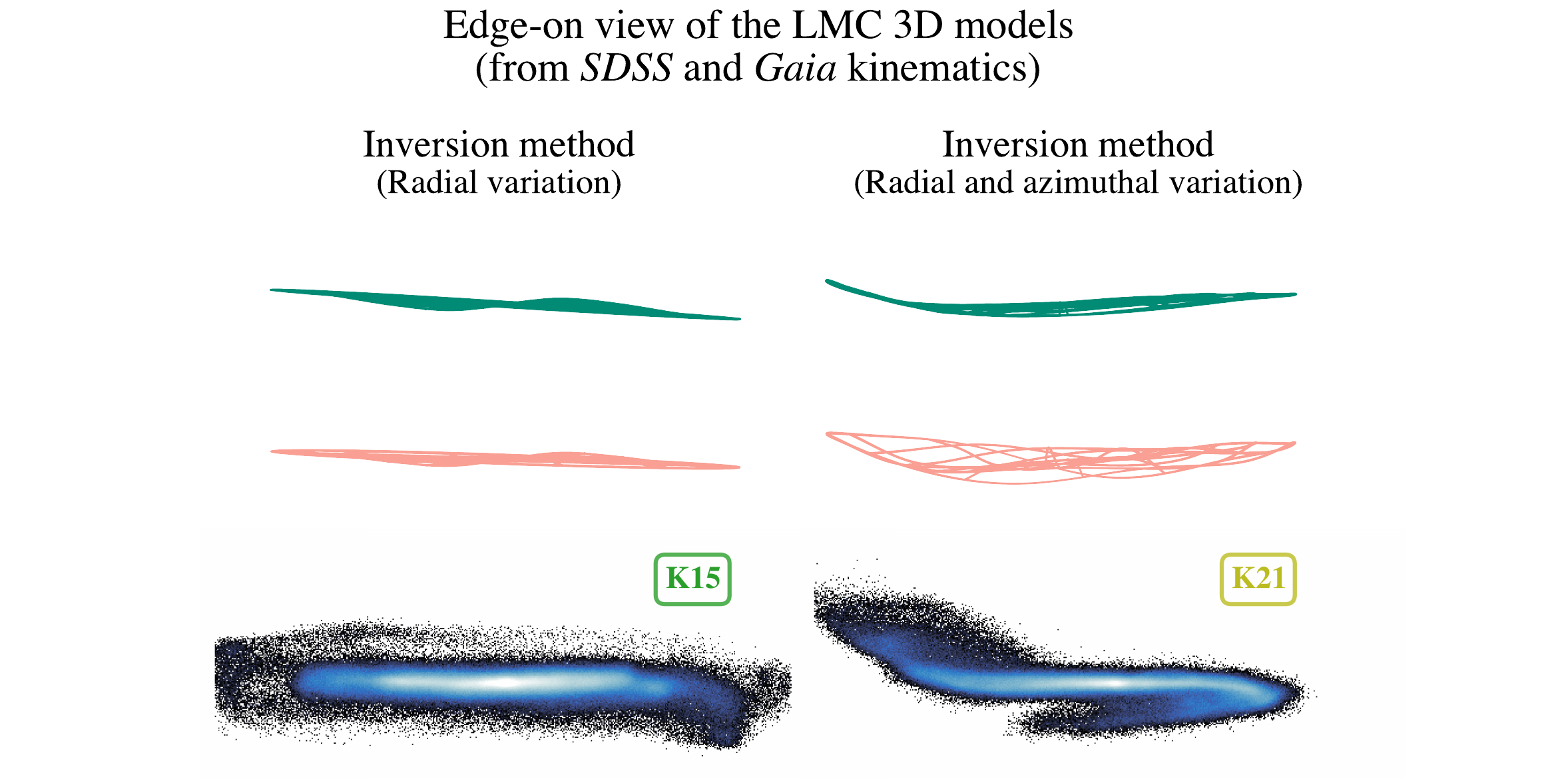}
    \caption{Edge-on view of LMC 3D models. First and second row left panels: Inversion method to best fit the inferred viewing angle variations with galactocentric radius $R'$ as a continuous surface. First and second right panels:  Inversion method to best fit the inferred viewing angle variations with both galactocentric radius $R'$ and azimuth $\phi$ as a continuous surface. To best highlight the tilted bar (left) and outer warp (right), the two 3D models are viewed from different angles, with a difference of $\Delta \phi = 70^\circ$ in the edge-on plane. Top panels (green): LMC Combined complete sample. Central panels (salmon): LMC Combined optimal sample. Bottom panels: For comparison, two $N$-body simulated LMC analogs from the \textsl{KRATOS} suite \citep{jimenez-arranz24b}, K15 (left) and K21 (right), respectively. These were selected visually for resembling the edge-on views above determined from the data.}
    \label{fig:3d_model_edge_on}
\end{figure*}

The 3D models of the LMC disc in   Fig.~\ref{fig:3d_model} all reveal that the inner regions are located in a different plane than the disc outskirts, given the  variation of the inclination with radius. The models in the right panels further reveal that there is warping in the outer disc, given the variation of the inclination with azimuth. To highlight these features more clearly, we show edge-on views of the 3D inversion-method models in 
Fig.~\ref{fig:3d_model_edge_on}, where edge-on is defined based on the viewing angles determined in Sect.~\ref{subsec:incorrect_plane}.
We adopt different viewing directions from within this edge-on plane to highlight separately the tilted bar (left, for the models with only radial variations) and the outer warp (right, for the models with both radial and azimuthal variations). 
The interpretation and origin of these findings is further discussed in Sect.~\ref{sec:discuss_kinematics}.

The case that these are real features and are not some observational or analysis artifact is supported by comparison with $N$-body simulations of the LMC's interaction with the SMC. Disturbances from the SMC can cause deviations from a flat mid-plane, with the details depending on the exact orbit and impact parameter \citep{besla12}. To illustrate this qualitatively, the bottom of Fig.~\ref{fig:3d_model_edge_on}  shows edge-on views of two simulated LMCs extracted from the 
\textsl{KRATOS} simulation suite. These were  selected visually from Fig.~3 of \citet{jimenez-arranz24b} for resembling the edge-on views determined here from the data. 
The similarities are striking, especially given that no attempt was made to quantitatively fit the simulations to these data.

\subsection{Robustness against position of the LMC center}
\label{subsec:LMC_center}

There has been ongoing discussion about the rotational center of the LMC for a long time; see for example Fig.5 of \cite{vandermarel&kallivayalil2014} or Fig. 10 of \citet{luri20}. This problem initially manifested with the photometric centre and the centre of rotation for the HI gas lying at different positions. \citet{vandermarel01} determined the photometric center to be $(81.28^\circ, -69.78^\circ)$, and also determined that the center of the outer isopleths, adjusted for viewing angle, to be $(82.25^\circ,-69.50^\circ)$. According to \citet{kim98}, the kinematic center of the HI gas disc is $(79.40^\circ, -69.03^\circ)$, while \citet{luks-rohlfs92} found that it is $(78.13^\circ, -69.00^\circ)$. A rotational center $(78.76^\circ \pm 0.52^\circ, -69.19^\circ \pm 0.25^\circ)$ that is near the center of rotation for the HI gas was later determined by \citet{vandermarel&kallivayalil2014} using Hubble Space Telescope proper motions in the LMC. However, it was noted that this was not consistent with the rotational center derived from studies of the line-of-sight velocity distribution in carbon stars \citep[][e.g. $81.91^\circ \pm 0.98^\circ, -69.87^\circ \pm 0.41^\circ$]{vdm02}. \citet{wan20} used \gaia DR2 proper motions, along with \textsl{SkyMapper} photometry by \citet{wolf18} to find the dynamic centres for carbon stars, RGB stars, and young stars to be $(80.90^\circ \pm 0.29^\circ, -68.74^\circ \pm 0.12^\circ)$, $(81.23^\circ \pm 0.04^\circ, -69.00^\circ \pm 0.02^\circ)$, and $(80.98^\circ \pm 0.07^\circ, -69.69^\circ \pm 0.02^\circ)$, respectively. A kinematical model for the LMC proper motions was fitted using \gaia eDR3 data by \citet{luri20}, yielding a center of $(81.01^\circ,-69.38^\circ)$ that is marginally closer to the photometric center than the HI center. More recently, \citet{kacharov24} using a 3D model of the LMC determined the LMC kinematic center to be $(80.29^\circ \pm 0.04^\circ , -69.25^\circ \pm 0.02 ^\circ)$.

In the present work we have used the photometric center, but as discussed, this may not coincide with the kinematic center. To assess the dependence of our results on the adopted LMC center, we repeated the analysis of Sect. \ref{sec:interpretation} using the most recent determination of the LMC kinematic center $(80.29^\circ \pm 0.04^\circ , -69.25^\circ \pm 0.02 ^\circ)$ by \citet{kacharov24}. Their center is offset by $\sim 0.55$ kpc from the photometric center. 

With this change, we find that all our key requests remain qualitatively unchanged, including the visual appearance of Figs. \ref{fig:samples_vertical_velocity} and \ref{fig:model_precession_nutation}-\ref{fig:disc_fit_vs_azimuth}. To get a sense of the quantitative differences, we note that the fits in Sect. \ref{subsec:precession} now yield $di/dt = 1.03 \pm 0.03 \text{ km s}^{-1} \text{ kpc}^{-1}$ ($di/dt = 1.32 \pm 0.02 \text{ km s}^{-1} \text{ kpc}^{-1}$) and $d\Omega/dt = 3.21 \pm 0.07 \text{ km s}^{-1} \text{ kpc}^{-1}$ ($d\Omega/dt = 4.33 \pm 0.06 \text{ km s}^{-1} \text{ kpc}^{-1}$) for the LMC Combined optimal (complete) sample. If we redo the analysis of Sect. \ref{subsec:incorrect_plane}, for the LMC Combined optimal (complete) sample we now recover a disc plane with inclination $i = 23.3^\circ\pm0.1^\circ$ ($i=19.5^\circ\pm0.1^\circ$) and line-of-nodes position angle $\Omega = 329.8^\circ \pm 0.3^\circ$ ($\Omega = 332.6^\circ \pm 0.2^\circ$). The differences compared to the results in those sections are comparable to the differences between the results from the complete and optimal samples. Hence, uncertainties in the position of the LMC center do not materially alter our results, or our understanding of their systematic uncertainties.

\section{Summary and discussion}
\label{sec:discussion}

The objective of this work has been to map the vertical structure of the LMC by combining data from  \textsl{SDSS} and \textsl{Gaia} surveys. Motivated by the new line-of-sight velocities from \textsl{SDSS}-V's Milky Way Mapper (MWM) for a large sample of LMC stars, our goal has been to provide a more comprehensive map of the LMC vertical (off-plane) velocity $v_{z'}$ than the one provided in \cite{jimenez-arranz23a} just using \gaia DR3 data (see their Fig. 17). 

The improvement for this work, with the use of the LMC Combined sample (combination of all line-of-sight velocities available, see Sect. \ref{sec:data}), in comparison to \cite{jimenez-arranz23a} comes in the increase of the number of stars by almost a factor of three -- see Table \ref{tabl:samples} for a summary of the numbers of stars per LMC sample. Perhaps even more importantly, the improvement also comes from having a better coverage of the LMC outskirts (see Fig. \ref{fig:samples_density}). We validated the reliability of our results regarding the balance between completeness and purity in our samples by conducting the same analysis on the NN complete and optimal LMC clean samples from \cite{jimenez-arranz23a} and finding almost identical outcomes, with minor variations. However, in cases where the two samples yield answers for a quantity of interest that differs by more than the statistical uncertainties, we discuss both LMC samples. The difference between the results for the two samples can then be used to assess any systematic uncertainties.

\subsection{The LMC's vertical velocity maps in context}

Initially, to create the LMC velocity maps (see Fig. \ref{fig:samples_vertical_velocity}), we used the inclination $(i = 34^\circ)$ and line-of-nodes position angle $(\Omega= 310^\circ)$ derived in \citet{luri20} and used in \citet{jimenez-arranz23a}. To do so, the authors fitted the LMC proper motion as a flat rotating disc, assuming circular orbits. However, instead of using \citet{luri20}'s systemic motion for both proper motion $(\mu_{\alpha*,0},\mu_{\delta,0})$ and line-of-sight velocity ($\mu_{z,0}$) values, we re-derived these quantities in this work, because asymmetries in the LMC are well-known \citep[e.g.][]{besla12,Yozin2014,pardy18} and there is variation in the kinematics of different populations \citep[see, for example, Table 1 of][]{luri20}.

With regard to the median vertical velocity maps of the different LMC clean samples (see Sect. \ref{subsec:vertical_velocity_maps_lmc_clean_samples} and Fig. \ref{fig:samples_vertical_velocity}), the Combined samples show a trend that is consistent with that reported in the inner regions by the \gaia data \citep{jimenez-arranz23a}, that expands in the outskirts. With half of the galaxy $(x' < 0)$ moving upward and the other half $(x' > 0)$ moving downward ---with respect to the assumed plane of the LMC disc---, the vertical velocity map displays a bimodal trend/dipole. In addition, there appears to be an increasing gradient of vertical velocity magnitude from the inner to the outer disc. As argued in \citet{jimenez-arranz23a}, there are (at least) four possible explanations for this: 1) the symmetry axis of the disc is not stationary in an inertial frame due to precession and/or nutation as a result of tidal torques \citep[e.g.][]{weinberg00,vdm02,olsen07,olsen11}; 2) an incorrect estimation of the disc viewing angles \citep[e.g.][]{jimenez-arranz23a}; 3) the existence of a galactic warp in the LMC's disc \citep[e.g.][]{Choi2018,saroon22}; 4) the LMC is not in dynamical equilibrium \citep[e.g.][]{belokurov19,Choi2022,jimenez-arranz24a,jimenez-arranz24b}; or a combination of some/all of these different effects. These hypothesis were tested in Sect. \ref{sec:interpretation}, and are discussed further in Sect.~\ref{sec:discuss_kinematics} below.

Interestingly, we found a ``blob'' of stars with negative vertical velocities in the inner arm's center, at roughly $(x',y') = (-1, -3)$ kpc  \citep[see also][]{jimenez-arranz23a}. This area overlaps with the supershell LMC~4, a high star-forming region of ionized gas approximately $\sim$1 kpc in size \citep{tenorio-tagle88,vallenari03,book08,dawson13b,fujii14,ou24}; this region has been argued to be the LMC's biggest bubble-like structure -- see Fig. \ref{fig:supershell_LMC4}. We believe this is the first observation of the supershell LMC~4 in the (vertical) kinematics space of resolved LMC stars.

We created the vertical velocity maps for the LMC Combined sample using the different evolutionary phases that are described in Sect. \ref{subsec:evolutionary_phases}, to assess how old/young populations manifest in the vertical velocity plane. Overall, the dipole seen in the vertical velocity maps of the Combined sample is observed for the various evolutionary phases of the LMC (see Fig. \ref{fig:evol_vertical_velocity}). This indicates that, while different stellar populations could be moving at different vertical velocities, the overall motion of the LMC disc is consistent across all evolutionary phases. We split the kinematic tracers into the following stellar types, following the convention from \citet{luri20}: Young (Young 1 + Young 2 + Young 3), blue loop (BL), asymptotic giant branch (AGB), and red giant branch (RGB) samples. 

Despite the Young sample being the least represented population, it perfectly traces the supershell LMC~4 observed in the Combined sample. The Young sample is mostly composed by Young 1 stars (see Fig. \ref{fig:samples_cmd}) which corresponds to very young main sequence with ages $\lesssim$ 50 Myr. These stars, as well as the BL sample (with ages $\lesssim$ 300 Myr), are largely observed in the supershell LMC~4 \citep[see Fig. 3 of][]{luri20}. However, this supershell is not seen in the AGB or RGB samples, that are formed by older stars ($\lesssim$ 1-2 Gyr). The fact that we see this kinematic feature in young populations and not old ones corroborates that it is connected to the recent star forming region; the ages of stars seen in this population could also serve as a proxy for setting an upper bound on the start of a burst in this region. According to \cite{dopita85}, star formation appears to have begun close to the centre about $\sim15$ Myr ago, which is consistent with the age of the stars in our Young sample but not in the BL sample. The authors also discovered three components, two of them being linked to gas shells ejected above and below the LMC disc plane at a velocity of $\sim36$ \kmsnospace. Our results suggest that we are possibly witnessing this phenomenon, and that it could have potentially begun at an earlier point in time.

Allowing for pure speculation, this kinematic feature could be caused by a potential crossing of the SMC with the LMC disc. This is because it is difficult to obtain high vertical velocity kinematics for a relatively small grouping of young stars in galaxy discs from secular evolution alone. Thus, we argue that our finding of the supershell LMC~4 with high negative vertical kinematics may be the result of one of these interactions between the LMC and SMC. If so, age dating the stellar populations in the LMC~4 supershell structure could help place further constraints on the timing of the interaction between these two sister galaxies. While this scenario is completely speculative at this point in time, it would be an interesting question to entertain in greater detail in future work using numerical simulations.

Additionally, we are able to determine the vertical velocity dispersion and confirm that the Young and BL samples, which represent the younger evolutionary phases, have a smaller velocity dispersion (of the order of $\sim$5 \kmsnospace) than the AGB and RGB samples, which represent the older evolutionary phases (see Fig.~\ref{fig:evol_vertical_velocity_dispersion}). This is to be expected, as older populations have had time to become kinematically hotter than younger populations. 

Finally, and for the first time ever, thanks to this wealth of data, we are able to present the one-dimensional vertical velocity profiles as a function of the azimuth, $\phi$, defined in the LMC in-plane $(x', y')$ Cartesian coordinate system (see Fig. \ref{fig:evol_1D_slices}). For the Combined sample we observe that the azimuthal wedges along the line-of-nodes ($x' \sim 0$) show a clear trend for the vertical velocity $v_{z'}$ close to the LMC center. The right panels of Fig. \ref{fig:samples_vertical_velocity} demonstrate a linear increase (decrease) in the vertical velocity as a function of the LMC galactocentric radius for those wedges. Conversely, the azimuthal wedges that are perpendicular to the line-of-nodes ($y' \sim 0$) exhibit flatter behaviour for all radii. Due to the limited coverage of RGB stars in the LMC centre, we cannot tell if the RGB stars show the observed linear trend of vertical velocity with respect to the galactocentric radius. In the LMC outskirts, however, it displays flattened curves and follows the same pattern as the LMC Combined sample. We can observe the same linear relationship close to the LMC centre for the AGB and BL stars because for these samples the LMC inner region has better coverage than for the outer regions.

\subsection{Interpretation of the LMC vertical kinematics}
\label{sec:discuss_kinematics}

After characterising the LMC vertical kinematic maps for the LMC Combined samples, in Sect. \ref{sec:interpretation} we set out to interpret them. To do so, we conducted three tests: 1) check if the vertical velocity dipole presented in Figs. \ref{fig:samples_vertical_velocity} and~\ref{fig:evol_vertical_velocity} could be explained by time variability in the disc symmetry axis (e.g., due to the rate of change of the LMC's disc viewing angles); 2) re-derive the LMC disc plane by minimising the vertical velocity $v_{z'}$ across the disc, as the LMC disc plane used in Sect. \ref{sec:vertical_velocity_maps} \citep[namely, the one fitted in][]{luri20} assumes that stars in the LMC disc are on circular orbits, which may not be the case; 3) re-estimate the values of the inclination angle, $i$, and line-of-nodes position angle, $\Omega$, using our data in different bins of galactocentric radius and azimuth. This is because the LMC's disc could be warped, and therefore does not require to be flat and characterized by constant viewing angles.

To undertake the first test, we re-derive a new estimate of the rate of change of the LMC disc viewing angles; namely, the inclination $di/dt$ and the line-of-nodes position angle $d\Omega/dt$, which results in a solid body rotation of the LMC disc. In this work we use a more direct method than the one used in \citet{vdm02, olsen07, olsen11}, where the authors estimated $di/dt$ tailored to a case without astrometric information, and assuming circular orbits to make up for the absence of full 3D velocity info. Since the line-of-sight velocity field does not provide any information on time variations in the position angle of the line of nodes, $d\Omega/dt$, they were unable to provide any estimate for that quantity.

Now, thanks to the astrometric and spectroscopic data provided by the \gaia and \textsl{SDSS} surveys, we have access to the vertical velocities for the LMC disc \citep[][and this work]{jimenez-arranz23a}, which allows us to estimate $di/dt$ and $d\Omega/dt$ using a more direct method. This method involves fitting the LMC vertical velocity under the assumption that the only signal originates from the change in the viewing angles of the LMC disc (see Eq. \ref{eq:model}). Using this method, we obtain $di/dt  = 0.76 \pm 0.03 \text{ km s}^{-1} \text{ kpc}^{-1}  = 45 \pm 2 ^\circ \text{ Gyr}^{-1}$ and $d\Omega/dt  = 2.83 \pm 0.07 \text{ km s}^{-1} \text{ kpc}^{-1}  = 166 \pm 4 ^\circ \text{ Gyr}^{-1}$ for the LMC Combined optimal sample. The results for the LMC Combined complete sample are somewhat different, but also significantly positive for both quantities (see Sect. \ref{subsec:precession}). However, for either sample, when we compare the data with the model (see Fig.~\ref{fig:model_precession_nutation}), the residual maps clearly have significant structure that cannot be attributed solely to time-variable viewing angles. Inclusion of non-zero 
$di/dt$ and $d\Omega/dt$ 
decreases the RMS for the $v_{z'}$ maps by only $\sim1\%$. Thus, other effects are likely responsible for the structure in the $v_{z'}$ maps, which could render 
our determinations of $di/dt$ and $d\Omega/dt$, under the assumptions of a solid rotating body, as meaningless. 

With this caveat in mind, it is nonetheless useful to compare our result for $di/dt$ to other estimates in the literature.\footnote{Other papers in the literature on this topic did not have access to astrometric data for individual stars, and hence could not provide an estimate on $d\Omega/dt$. } First, \citet{vdm02} provided a value of $di/dt = -0.37 \pm 0.22$ mas yr$^{-1} = -103 \pm 61^\circ$ Gyr$^{-1}$ using line-of-sight measurements for 1,041 carbon star. Then, \citet{olsen07} and \citet{olsen11} with new observations of carbon stars and red supergiants provided an estimates of $di/dt = 0.001$ mas yr$^{-1}$ and $di/dt = -184 \pm 81^\circ$ Gyr$^{-1}$, respectively. Our results disagree with these values at the $\sim 3\sigma$ level. However, this is not entirely surprising given the caveats discussed regarding the interpretation of $v_z'$ maps. 

Theoretical estimates of the rates of change of the LMC disc viewing angles with numerical models \citep[e.g.][]{weinberg00} are left for future work. This could be addressed with the \texttt{KRATOS} \citep{jimenez-arranz23b} simulations, for example. \texttt{KRATOS} is a comprehensive suite of 28 pure N-body simulations of isolated and interacting LMC-like galaxies devoted to study the formation of substructures in their disc after the interaction with an SMC-mass galaxy, or by identifying LMC analogues in cosmological simulations \citep[e.g.][]{boylan-kolchin11,sales11,santos-santos21,buch24}. We reserve this exploration for future work.

After discarding any time-variability in the LMC disc viewing angles as the main reason for the dipole and structure in the vertical velocity maps, we considered alternative explanations. Since the analysis above was conducted under the assumption that the LMC disc plane fitted in \citet{luri20} was the actual LMC disc plane, this could yield a biased answer because of the assumption of circular orbits. To test whether this assumption is indeed biasing our results, we re-derived the orientation of the LMC disc plane by minimising the RMS vertical velocity $v_{z'}$ across the disc of the LMC (see Sect. \ref{subsec:incorrect_plane}). For the LMC Combined optimal (complete) sample, we recover a disc plane with inclination $i =23.6^\circ\pm0.1^\circ$ ($i=20.0^\circ\pm0.1^\circ$) and line-of-nodes position angle $\Omega = 326.9^\circ \pm 0.3^\circ$ ($\Omega = 330.0^\circ \pm 0.2^\circ$). Compared to the LMC disc plane fitting of \citet{luri20}, with $i = 34^\circ$ and $ \Omega=310^\circ$, the LMC disc plane inclination and line-of-nodes position angle that we recover for the LMC Combined optimal (complete) sample differ by $\delta i \sim 10^\circ$ ($\delta i \sim 14^\circ$) and $\delta \Omega \sim 17^\circ$ ($\delta \Omega \sim 20^\circ$). The values we obtained are consistent with the low-end of the wide range of estimates found in the literature \citep[e.g.][]{vdm09,vandermarel&kallivayalil2014,Haschke&Grebel&Duffau2012,ripepi22,kacharov24}.

When analysing the new kinematic maps with the new fitted values of $i$ and $\Omega$ in this work (see Fig. \ref{fig:maps_correct_plane}), we observe that: $1)$ the LMC bar major axis becomes less aligned with the line-of-nodes due to the change of plane -- note that there is no physical reason for both axes being aligned; $2)$ the LMC's bar bimodal vertical velocity trend vanishes, leaving behind a nearly null median vertical velocity; $3)$ there is a sign change in the disc's vertical velocity bimodal trend; $4)$ the supershell LMC~4 continues to stand out from the background, albeit not as much as with the LMC disc plane fitted in \cite{luri20}. We found that a different disc orientation (although relatively similar) than the one suggested by \citet{luri20} minimizes the overall vertical velocities. However, even with that new orientation a clear dipole pattern remains visible in the vertical velocity maps.

We therefore undertook a third test to investigate whether this may be due to warps or twists in the LMC disc plane. To do so, we repeated the best fit disc plane minimisation of Sect. \ref{subsec:incorrect_plane} for different annular rings in the $(x',y')$ plane -- defined by the inclination angle, $i$, and line-of-nodes position angle, $\Omega$, of \cite{luri20} -- to quantify the change of $(i,\Omega)$ as function of the LMC galactocentric radial position $R'$ (see Fig. \ref{fig:disc_fit_vs_radius} left panels). We observed that in both samples the inclination is low ($i \sim 10^\circ$) at the inner regions ($R' \sim 0$ kpc) and grows up to values of $i \sim 25^\circ$ for the outer disc ($R' \gtrsim 3.5$ kpc). The line-of-nodes position angle in both samples is high ($\Omega \sim 335^\circ - 350^\circ$) at the inner regions ($R' \sim 0$ kpc), decreases to values of $\Omega \sim 320^\circ - 325^\circ$ at approximately $R' \sim 2.5$ kpc, and then increases once more to values of $\Omega \sim 335^\circ$ in the outer disc ($R' \gtrsim 3.5$ kpc). For both inclination and line-of-nodes position angle, the LMC Combined complete and optimal samples have a small and almost constant offset from each other. Regardless, the important result from these findings is that the LMC disc is not flat. 

Other observational studies have also found radial variations in the disc viewing angles in support of warps and twists in the LMC disc, but the results are not always mutually consistent. 
\citet{vdmcioni01} performed a photometric study of tracer brightnesses as function of position on the sky. They found a decrease in inclination from $i \sim 40^\circ$ just outside the bar to $i \sim 30^\circ$ for the outer disc, as well as a gradual decrease in the position angle of the line-of-nodes. The declining inclination trend in the outer regions may be consistent with our findings, albeit with a smaller amplitude. However, we find an increase in the line-of-nodes position angle with at large radii, instead of a decrease. Compared to the ``global'' fit presented in this work ($i \sim 20 - 23^\circ$), the inclination derived in \citet{vdmcioni01} is larger ($i \sim 35^\circ$), and the line-of-nodes position angle lower ($\Omega \sim ~303^{\circ}$), both being more comparable to the values from \citet{luri20}. The photometric study of \citet{Choi2018} ignored the innermost regions of the LMC disc. Beyond that, in contrast to our findings and more similar to the photometric results of \citet{vdmcioni01}, they inferred that the inclination decreases as a function of radius, beginning at $i \sim 40^\circ$. However, more similar to our findings, the authors found a decreasing trend as function of radius for the line-of-nodes position angle beyond the bar, as shown in Fig~\ref{fig:disc_fit_vs_radius}.

We also quantified the change of $(i,\Omega)$ as a function of both the LMC azimuth $\phi$ and galactocentric radial position $R'$ by repeating the best disc plane fitting in annular wedges (see Fig. \ref{fig:disc_fit_vs_azimuth}). In agreement to the annular fits, both the LMC Combined complete and optimal sample display a fairly homogeneous and isotropic central part of the disc with low inclination $i$ and high line-of-nodes position angle $\Omega$. However, we observe a quadrupolar pattern for both angles $(i,\Omega)$ in the outer regions, with a significant variation as a function of the azimuth $\phi$. This is a clear sign of a warp in the LMC disc plane.

The inferred spatially variable viewing angles have enabled us to construct three-dimensional models of the shape of the LMC disc plane, using both the LMC Combined complete and optimal samples (see Fig. \ref{fig:3d_model}). For this we have used both traditional tilted-ring models, as well as a new inversion method described in the Appendix. The 3D models show that the inner regions are located in a different plane than the disc outskirts, given the 
difference in inclination found between the inner and outer regions of the LMC disc. This could be due to an inclination of the bar relative to the disc plane \citep[e.g.][]{besla12,Choi2018}. Simulations of the LMC-SMC interaction by \citet{besla12} indeed found a $5^\circ - 15^\circ$ tilted bar with regard to the LMC disc, if
the SMC recently passed through the LMC disc. 
\citet{Choi2018} postulated that such a tilt was consistent with the observed spatial variations in the magnitude of red clump stars. \citet{Choi2022} reported additional evidence for a recent SMC collision with a small impact parameter.
Our results provide further support for these findings.
However, it must be noted that other effects associated to disequilibrium \citep[e.g.][]{belokurov19,Choi2022,jimenez-arranz24a,jimenez-arranz24b} may play a role as well. Previous authors have found evidence for a 
warp in the outer regions of the LMC disc \citep[e.g.][]{vdm02,Olsen2002,nikolaev04,Choi2018,ripepi22}. Our 3D models support this, given the variation of the inclination with azimuth found in the disc. The appearance of an inclined bar could merely be a natural extension of this warp, instead of a separate physical effect.

\section{Conclusions}
\label{sec:conclusions}

The aim of this work has been to map the vertical structure of the LMC by combining data from  \textsl{SDSS} and \textsl{Gaia} surveys.  This was motivated by the new line-of-sight velocities from \textsl{SDSS}-V's Milky Way Mapper (\textsl{MWM}) for a large sample of LMC stars. These enabled the construction of a more comprehensive map of the LMC vertical (off-plane) velocity $v_{z'}$ than the one provided in \cite{jimenez-arranz23a} using only \gaia DR3 data (see their Fig. 17). With this improved map, we were able to perform a comprehensive investigation of the 3D shape of the LMC disc plane, and of the orientation of its symmetry axis and any time-variability therein. The main results shown in this paper can be summarised as:

\begin{itemize}
    \item We increased the number of LMC stars with line-of-sight velocity information by almost a factor of three compared to \citet{jimenez-arranz23a}. Even more importantly, we have better coverage of the LMC disc outskirts. With these data, we cover a projected radius of $R\gtrsim5$ kpc (see Fig.~\ref{fig:samples_density}).
   
    \bigskip\item We provided the first identification of the supershell LMC~4 in the (vertical) kinematics space of resolved LMC stars (see Figs.~\ref{fig:samples_vertical_velocity}-~\ref{fig:evol_vertical_velocity_dispersion}). When examining various stellar populations, we observed that it is only discernible for stars younger than $\lesssim 300$ Myr (see Fig.~\ref{fig:evol_vertical_velocity}). We speculate that this might be an aftermath of the recent LMC-SMC interaction, which will be investigated in a future work. 
    
    \bigskip\item In agreement to \citet{jimenez-arranz23a}, but expanding the trend in the outskirts of the LMC,  half of the galaxy moves upward and the other half moves downward with respect to the fitted plane of the LMC disc provided by \citet[][see Figs.~\ref{fig:samples_vertical_velocity} and \ref{fig:evol_1D_slices}]{luri20}. This is also the case when examining the profile for stars at different evolutionary stages. 
    
    \bigskip\item We interpreted this bimodal dipole/trend in the LMC vertical velocity map with three distinct possibilities: 1) time-variation in the orientation of the disc symmetry axis, (e.g.~due to precession or nutation); 2) use of an incorrect LMC disc plane, and; 3) the presence of warps and twists in the LMC disc.
    
    \bigskip\item 1) We fitted for  time-variation in the disc viewing angles, and obtain non-zero results, e.g., $di/dt = 0.76 \pm 0.03 \text{ km s}^{-1} \text{ kpc}^{-1}  = 45 \pm 2 ^\circ \text{ Gyr}^{-1}$ and $d\Omega/dt  = 2.83 \pm 0.07 \text{ km s}^{-1} \text{ kpc}^{-1}  = 166 \pm 4 ^\circ \text{ Gyr}^{-1}$ for the LMC Combined optimal sample (see Fig.~\ref{fig:model_precession_nutation}). However, these results may be spurious, since most of the structure in the vertical velocity maps remains even when these fitted values are included. So time variation in the orientation of the disc symmetry axis does not explain the inferred velocity dipole.
    
    \bigskip\item 2) We re-derived the orientation of the LMC disc plane by minimising the RMS vertical velocity $v_{z'}$ across the disc of the LMC. For the LMC Combined optimal sample, we recover a disc plane with inclination $i = 23.6^\circ\pm0.1^\circ$ and line-of-nodes position angle $\Omega = 326.9^\circ \pm 0.3^\circ$ (see Fig.~\ref{fig:maps_correct_plane}). The results for the LMC Combined complete sample differ by $\sim~3^{\circ}$, which provides an indication of the systematic uncertainty in these determinations. Changes in the adopted position of the LMC center can affect the results at similar levels.
    
    \bigskip\item 3) We repeated the plane fitting for different annular rings in the LMC disc (see Fig.~\ref{fig:disc_fit_vs_radius}). This implies a difference in inclination between the inner and outer regions of the LMC disc. This is qualitatively consistent with a $5^\circ - 15^\circ$ tilted bar with regard to the LMC disc as seen in simulations by \citet{besla12}, and previously supported observationally in  \citet{Choi2018}. We also quantified the change of the viewing angles as a function of both azimuth and galactocentric radius (see Fig.~\ref{fig:disc_fit_vs_azimuth}). This yield a quadrupolar pattern for both viewing angles in the outer disc. The variations as a function of  azimuth imply the presence of a warp in the outer LMC disc plane. 
    
    \bigskip\item We presented a new inversion method to use the results from the aforementioned analyses to construct 3D representations of the shape of the LMC disc plane (see Figs.~\ref{fig:3d_model} and \ref{fig:3d_model_edge_on}). We presented the results from this method, and compared them to more traditional tilted-ring representations. The 3D representations reveal that the LMC disc is not a flat plane in equilibrium, but that the central bar region is tilted relative to a warped outer disc. These provide further evidence for  perturbations caused by interaction with the SMC.

\end{itemize}

In summary, the currently available \textsl{SDSS} and \gaia dataset has proven suitable for studies of the LMC out-of-plane kinematics and 3D structure. Future releases of surveys like \textsl{SDSS}, \gaia (DR4), and \textsl{4MOST} may allow for a revisit of this work once more line-of-sight velocities are available in the LMC.

\section*{Acknowledgements}

We are grateful to M. Romero-Gómez, X. Luri, K. V. Johnston, E. Cunningham, H. Rathore, and J. G. Fernández-Trincado for fruitful discussions.  OJA acknowledges funding from ``Swedish National Space Agency 2023-00154 David Hobbs The GaiaNIR Mission'' and ``Swedish National Space Agency 2023-00137 David Hobbs The Extended Gaia Mission''. DH acknowledges support from the Flatiron Institute, funded by the Simons Foundation. CL acknowledges funding from the European Research Council (ERC) under the European Union’s Horizon 2020 research and innovation programme (grant agreement No. 852839).

Funding for the Sloan Digital Sky Survey V has been provided by the Alfred P. Sloan Foundation, the Heising-Simons Foundation, the National Science Foundation, and the Participating Institutions. SDSS acknowledges support and resources from the Center for High-Performance Computing at the University of Utah. SDSS telescopes are located at Apache Point Observatory, funded by the Astrophysical Research Consortium and operated by New Mexico State University, and at Las Campanas Observatory, operated by the Carnegie Institution for Science. The SDSS web site is \url{www.sdss.org}.

SDSS is managed by the Astrophysical Research Consortium for the Participating Institutions of the SDSS Collaboration, including the Carnegie Institution for Science, Chilean National Time Allocation Committee (CNTAC) ratified researchers, Caltech, the Gotham Participation Group, Harvard University, Heidelberg University, The Flatiron Institute, The Johns Hopkins University, L'Ecole polytechnique f\'{e}d\'{e}rale de Lausanne (EPFL), Leibniz-Institut f\"{u}r Astrophysik Potsdam (AIP), Max-Planck-Institut f\"{u}r Astronomie (MPIA Heidelberg), Max-Planck-Institut f\"{u}r Extraterrestrische Physik (MPE), Nanjing University, National Astronomical Observatories of China (NAOC), New Mexico State University, The Ohio State University, Pennsylvania State University, Smithsonian Astrophysical Observatory, Space Telescope Science Institute (STScI), the Stellar Astrophysics Participation Group, Universidad Nacional Aut\'{o}noma de M\'{e}xico, University of Arizona, University of Colorado Boulder, University of Illinois at Urbana-Champaign, University of Toronto, University of Utah, University of Virginia, Yale University, and Yunnan University.

This work has made use of data from the European Space Agency (ESA) mission {\it Gaia} (\url{https://www.cosmos.esa.int/gaia}), processed by the {\it Gaia} Data Processing and Analysis Consortium (DPAC, \url{https://www.cosmos.esa.int/web/gaia/dpac/consortium}). Funding for the DPAC has been provided by national institutions, in particular the institutions participating in the {\it Gaia} Multilateral Agreement.

\bibliographystyle{aa}
\bibliography{mylmcbib} 

\appendix

\section{Determination of 3D LMC Shape}

Figure~\ref{fig:disc_fit_vs_azimuth} shows determinations of the LMC viewing angles $i$ and
$\Omega$ on a polar grid in the $(x',y')$ plane. The grid consists of segments that have their centres at radii $R'_l = \Delta R' (l-{1\over 2}) $ for $l = 1, \ldots L$ and $\phi'_j = \Delta \phi' (m-{1\over 2})$ for $m = 1, \ldots M$. Here $R'$ is the cylindrical radius in the
$(x',y')$ plane, and $\phi'$ the azimuthal angle in that plane measured counter-clockwise from the $x'$-axis. In practice, we take $\Delta R' = 1$ kpc with $L=6$ and $\Delta R' = 360^{\circ}/M$, with $M=12$. For the discussion below we group the innermost $M$ azimuthal segments
into a single filled circle of radius $\Delta R'$, centred on
$R'=0$. There are then a total of $N_{\rm seg} = 1 + (L-1) \> M$
segments.

On each individual segment $n=1,\ldots,N_{\rm seg}$ the LMC disc is
described by a plane of the form $z' = a_n (x'- x'_{{\rm cen},n}) +
b_n (y'- y'_{{\rm cen},n}) + z'_{{\rm cen},n}$. Here $(x'_{{\rm
cen},n}, y'_{{\rm cen},n}, z'_{{\rm cen},n})$ is the center of the segment in the 3D $(x',y',z')$ coordinate space. The normal vector to this plane is $(a_n,b_n,-1)$. This normal vector is fully defined by the viewing angles $i_n$ and $\Omega_n$ for the given segment, so that
the $a_n$ and $b_n$ are known through the transformation equations given in \cite{vandermarel01} and \cite{vdm02}. However, the $z'_{{\rm cen},n}$ for $n=1,\ldots,N_{\rm seg}$ are unknown and need to be determined to fully describe the three-dimensional shape of the LMC disc.

To determine the $z'_{{\rm cen},n}$ we enforce the constraint that the disc be continuous at the segment boundaries. The boundaries between adjacent segments within the same annulus are centred on the points with polar coordinates $R'_l$ as defined above for $l = 2, \ldots L$ and $\phi'_k = k \Delta \phi'$ for $k = 1, \ldots M$. The boundaries between adjacent segments in adjacent annuli are centred on the
points with polar coordinates $R'_j = j \Delta R'$ for $j = 1, \ldots L-1$ and $\phi'_m$ as define above for $m = 1, \ldots M$. The are a total of of $N_{\rm bound} = 2 \> (L-1) \>M$ of such boundary centres. To lowest order, there are about twice as many boundary centres as segment centres.

For each boundary center $i = 1,\ldots, N_{\rm bound}$ there are two adjacent segments $n_1$ and $n_2$ that touch, with $n_1$ and $n_2$ in the range $1, \ldots, N_{\rm seg}$. To get any two adjacent segments to yield the same value of $z'$ at their boundary center, the following equations must be satisfied for $i = 1,\ldots, N_{\rm bound}$:
\begin{eqnarray}
  z'_{{\rm cen},n1} - z'_{{\rm cen},n2} = a_{n2} (x'_i - x'_{{\rm cen},n2}) - a_{n1} (x'_i - x'_{{\rm cen},n1}) + \nonumber \\ b_{n2} (y'_i - y'_{{\rm cen},n2}) - b_{n1} (y'_i - y'_{{\rm cen},n1})     .
\label{eq:continuity}		      
\end{eqnarray}
The right-hand side is known for each $i$ and its corresponding $n_1$ and $n_2$, whereas the left-hand side contains two of the unknowns. This set of equations can be expressed as a matrix equation of the form ${\bf A} {\bf z'}= {\bf c}$. Here ${\bf A}$ is a sparse $N_{\rm bound} \times N_{\rm seg}$ matrix with only elements $+1$, $-1$, and
$0$; ${\bf c}$ is the $N_{\rm bound}$ vector composed of the right-hand sides of equation~(\ref{eq:continuity}); and ${\bf z'}$ is the $N_{\rm
seg}$ vector composed of the unknown $z'_{{\rm cen},n}$ for
$n=1,\ldots,N_{\rm seg}$. 

The aforementioned matrix equation ${\bf A} {\bf z'}= {\bf c}$ is an
overdetermined equation since $N_{\rm bound} > N_{\rm seg}$. It
generally has a well-defined solution in a least-squares sense, apart from the one-dimensional degeneracy that for any solution vector $z'_{{\rm cen},n}$, the addition of the same constant to all the vector elements also yields a solution vector. In other words, while we can determine the 3D shape of the LMC disc given $i$ and $\Omega$ on a polar grid, we cannot determine its distance. We solve this degeneracy by forcing the average of the elements of $z'_{{\rm cen},n}$ to be zero. The solution can then be determined using singular-value decomposition of the matrix ${\bf A}$, which was
implemented in {\tt FORTRAN} using subroutines from \citet{press92}.

After having determined the $z'_{{\rm cen},n}$ on the polar grid for $n=1,\ldots,N_{\rm seg}$, we use bi-cubic interpolation on the polar grid \citep[also using subroutines from][]{press92}  to allow evaluation of the surface $z'(x',y')$ for arbitrary $(x',y')$. This yields a continuous representation of the LMC disc that best matches
the $(i,\Omega)_n$ for $n=1,\ldots,N_{\rm seg}$ determined on a polar grid as described in Sect.~\ref{subsec:warps}. Once we have the bi-cubic interpolation, we can numerically determine the normal vector to the surface at the segment centres. Using the transformation equations
given in \cite{vandermarel01} and \cite{vdm02}, this yields the $(i,\Omega)_{{\rm fit},n}$ of the fitted surface.

The fitted $(i,\Omega)_{{\rm fit},n}$ do {\it not} generally reproduce the input $(i,\Omega)_n$ exactly, but they provide the best possible match with a {\it continuous} disc surface in the sense described above. As
discussed in Sect.~\ref{subsec:warps}, for our real LMC data there are significant residuals, but the
fitted $(i,\Omega)_{\rm fit}$ maps do capture the qualitative essence
of the input $(i,\Omega)$ maps.

The entire numerical implementation of the aforementioned algorithm was tested and validated by having it successfully recover known analytical surfaces from gridded 
$(i,\Omega)_{{\rm fit},n}$ pseudo-input data generated from those surfaces.

The formalism described above can also be used to generate 3D
renditions of the LMC disc for the situation of Sect.~\ref{subsec:incorrect_plane}, in which
the LMC viewing angles $i$ and $\Omega$ were fitted over annuli. In
this case we use the same polar grid as described above, with
$M=12$. But we set the input viewing angles to be the same for all segments within the same annulus. In this case we find that to maintain continuity, the fitted $(i,\Omega)_{\rm fit}$ do {\it not} have the same values for all segments within the same annulus. This is because by its very nature, a tilted-ring model is a disjunct, and not a continuous representation of a disc.


\label{lastpage}

\end{document}